\documentclass[namedreferences]{SolarPhysics}
\usepackage[optionalrh,natbib]{spr-sola-addons} % For Solar Physics
\usepackage{graphicx}                    % For eps figures, newer & more powerfull
\usepackage{amssymb}                     % useful mathematical symbols
\usepackage{color}                       % For color text: \color command
\usepackage{url}                         % For breaking URLs easily through lines
\newcommand{\arcsec}{^{\shortparallel}}

\newcommand{\aap}{    {\it Astron. Astrophys.}}

\newcommand{\apj}{    {\it Astrophys. J.}}
\newcommand{\apjl}{   {\it Astrophys. J.}}

\newcommand{\grl}{    {\it Geophys. Res. Lett.}}

\newcommand{\jgr}{    {\it J. Geophys. Res.}}

\newcommand{\pasp}{   {\it Pub. Astron. Soc. Pac.}}
\newcommand{\pasj}{   {\it Pub. Astron. Soc. Japan}}

\newcommand{\solphys}{{\it Solar Phys.}}

\newcommand{\planss}{ {\it Plan. Space Sci.}}

\renewcommand{\apjl}{   {\it Astrophys. J. Lett.}}

\renewcommand{\pasp}{   {\it Publ. Astron. Soc. Pacific}}
\renewcommand{\pasj}{   {\it Publ. Astron. Soc. Japan}}
\renewcommand{\planss}{ {\it Planet. Space Sci.}}
\begin{document}

\begin{article}

\begin{opening}

\title{Coronal Dimmings and the Early Phase of a CME Observed with STEREO and {\it Hinode}/EIS}

%%%%%%%%%%%%%%%%%%%%%%%%%%%%%%%%%%%%%%%%%%%%%%%%%%%
%% Authors Names
%
\author{C.~\surname{Miklenic}$^{1}$\sep
       A.M.~\surname{Veronig}$^{2}$\sep
        M.~\surname{Temmer}$^{1,2}$\sep
       C.~\surname{M\"ostl}$^{1,2}$\sep
       H.~K.~Biernat$^{1}$}

%%%%%%%%%%%%%%%%%%%%%%%%%%%%%%%%%%%%%%%%%%%%%%%%%%%
%% Runningheads
%
%\runningauthor{Miklenic et al.}
%\runningtitle{Coronal dimmings and the early phase of a CME}

%%%%%%%%%%%%%%%%%%%%%%%%%%%%%%%%%%%%%%%%%%%%%%%%%%%
%% Affilations
%
  \institute{$^{1}$ Space Research Institute, Austrian Academy of Sciences,
   Schmiedlstra\ss e 6, A-8042 Graz, Austria        \\
             $^{2}$ Kanzelh\"ohe Observatory-IGAM, Institute of Physics, University of Graz,
Universit\"atsplatz 5, A-8010 Graz, Austria \\ email: \url{asv@igam.uni-graz.at}
             }

%%%%%%%%%%%%%%%%%%%%%%%%%%%%%%%%%%%%%%%%%%%%%%%%%%%
%%% Abstract
\begin{abstract}
We investigate the early phase of the 13 February 2009 coronal mass ejection (CME). Observations with the twin STEREO spacecraft in quadrature allow us to compare for the first time in one and the same event the temporal evolution of coronal EUV dimmings, observed simultaneously on-disk and above-the-limb. We find that these dimmings are synchronized and appear during the impulsive acceleration phase of the CME, with the highest EUV intensity drop occurring a few minutes after the maximum CME acceleration. During the propagation phase two confined, bipolar dimming regions, appearing near the footpoints of a pre-flare sigmoid structure, show an apparent migration away from the site of the CME-associated flare. Additionally, they rotate around the `center' of the flare site, {\it i.e.}, the configuration of the dimmings exhibits the same 'sheared-to-potential' evolution as the postflare loops. We conclude that the motion pattern of the twin dimmings reflects not only the eruption of the flux rope, but also the ensuing stretching of the overlying arcade. Finally, we find that: (1) the global-scale dimmings, expanding from the source region of the eruption, propagate with a speed similar to that of the leaving CME front; (2) the mass loss occurs mainly during the period of strongest CME acceleration.  Two hours after the eruption \textit{Hinode}/EIS observations show no substantial plasma outflow, originating from the `open' field twin dimming regions.
\end{abstract}%%%%%%%%%%%%%%%%%%%%%%%%%%%%%%%%%%%%%%%%%%%%%%%%%%%
%% Keywords
%
\end{opening}
%-------------------------------------------------

%%%%%%%%%%%%%%%%%%%%%%%%%%%%%%%%%%%%%%%%%%%%%%%%%%%
%% Sections
%
%%%%%%%%%

\section{Introduction}

Coronal dimmings are regions in the solar corona that undergo an abrupt intensity drop roughly co-temporal with the launch of a coronal mass ejection (CME). They are observable in soft X-rays \citep[SXR;][]{sterling97} and most conspicuously in the extreme UV \citep[EUV; {\it e.g.},][]{zarro99,thomson00} both on the solar disk and above the limb.

Global-scale dimmings originate from the source region of the eruption and may expand across almost the entire solar disk \citep[{\it e.g.},][]{zhukov04,mandrini07}. They follow the nearly spherically propagating fronts of global coronal waves \citep[Moreton waves and EIT waves,][]{moreton60,moses97,thompson98}. On the other hand, small-scale twin dimmings, which typically exhibit a more dramatic intensity decrease, occur in confined areas near the ends of a pre-flare, S-shaped sigmoid structure \citep{sterling97}.

The generally accepted physical interpretation of coronal dimmings is that they reflect the strong density depletion of the inner corona due to the expansion or `opening' of magnetic field lines in the wake of a CME.
Two major lines of evidence support this view: (1) Measurements of Doppler shifts in emission lines formed in the lower transition region and corona revealed material outflows co-spatial with small-scale dimming regions, with outflow velocities of several tens of $\rm{km\,s^{-1}}$ \citep{harra01,harra07,jin09}. Inside global-scale dimmings mass outflows with velocities of $\approx \rm{100\,km\,s^{-1}}$ were found \citep{harra01}. (2) Based on the idea that due to mass conservation the mass accumulated in the leading edge of the CME has to be balanced by the density depletion in the dimming region, coronal dimmings have been used to estimate CME masses \citep[{\it e.g.},][]{harrison00,harrison03,aschwanden09b,aschwanden09a,jin09}. In some cases, the consistency with CME masses based on white-light measurements was good \citep[{\it e.g.},][]{harrison03,aschwanden09b}, in other cases, the CME mass based on EUV observations was 30--70\% smaller than the white-light estimate \citep[{\it e.g.}, ][]{harrison00,jin09}. \citet{jin09} attributed this discrepancy partly to transition region outflows, which refill the dimming regions. Nevertheless, these studies strengthen the link between coronal dimmings and CMEs. Therefore, the investigation of the dimming process is considered as a powerful diagnostic of the early phase of a CME, especially with regard to the physics of CME onsets \citep{harrison03}. This phase cannot be examined using white-light observations, since the occulting disk of the coronagraph obscures both the Sun and the lower layers of the corona, where the CME is usually launched and accelerated \citep{temmer08,temmer10}.

Previous studies on coronal EUV dimmings addressed their temporal and spatial relationship to the associated CME, the amount of the EUV intensity drop as well as the timescale and duration of the dimmings \citep[{\it e.g.},][]{sterling97,aschwanden99,mcintosh07,bewsher08}. \citet{mcintosh07} examined both the temporal evolution of the size of particular dimming areas and the apparent motion of these regions. Due to observational limitations, however, in all of these studies a particular dimming event was observed either on the solar disk or above the limb, depending on the particular line of sight from the observing instrument to the source region of the eruption. Therefore, it was impossible to examine in one and the same event the temporal evolution of coronal dimmings from different vantage points.

With the launch of the twin STEREO spacecraft \citep[{\it Solar-Terrestrial Relations Observatory},][]{kaiser08} these observational limitations are temporarily over, and more importantly, the Sun was cooperative enough to produce a CME/flare event when STEREO Ahead (STA) and STEREO Behind (STB) were in quadrature on 13 February 2009. The resulting unprecedented data set allows us to investigate the coronal EUV dimming process in this event simultaneously both on the solar disk and above the limb. Due to the line-of-sight integration of the optically thin emission, the combination of on-disk and above-the-limb observations is particularly instructive. We compare both dimming observations and relate them to the kinematics of the associated CME. Finally, we combine EUV observations obtained from both the STEREO and the \textit{Hinode} spacecraft \citep{kosugi07} to examine the relationship between dimming locations and mass flows.

In Section~\ref{sec:dobs} we describe the observations as well as the data sets used and the data reduction procedures applied. The analysis and results are presented in Sections~\ref{sec:analysis} and \ref{sec:results}. In Section~\ref{sec:discussion} we discuss the results.

\section{Observations and Data Reduction}\label{sec:dobs}

The 13 February 2009 CME/flare event occurred in NOAA active region AR 11012. At this time, the separation angle of the two STEREO spacecraft was $91^\circ$.
STA observed the event above the solar limb, while STB observed it near the center of the solar disk.
 The CME had a typical three-part structure, {\it i.e.}, a bright front, a dark cavity, and a bright core. The flare (GOES class B2.4) started around 05:30~UT (GOES maximum at 05:47 UT). The event was associated with a global coronal wave observed in EUV \citep{cohen09,kienreich09,patsourakos09}.

Although in this paper we focus on the investigation of the early phase of the CME, it is interesting to note that the Heliospheric Imagers \citep[HI,][]{eyl09} onboard STA tracked the CME's leading edge up to the distance of about 1~AU. In addition, the STB/IMPACT \citep{luh08} and PLASTIC \citep{gal08} instruments registered a weak shock on 18 February 2009, 10:00~UT, followed by the complex magnetic field and bulk plasma signatures typically associated with an interplanetary coronal mass ejection (ICME). A possible encounter of the ICME with Venus was observed with the Venus Express magnetometer \citep{zha06}. A detailed study on the ICME using the STA/HI observations along with the in situ measurements provided by STB and the Venus Express is underway and will be presented elsewhere \citep{moestl11}.

\subsection{STEREO Data}
From the STA and STB spacecraft we use EUV 171 and 195~{\AA} filtergrams, provided by the Extreme Ultraviolet Imager \citep[EUVI;][]{wuelser04,howard08}, to investigate the dimming process both on the solar disk and above the limb. In addition, we use the STA/EUVI 171~{\AA} filter in combination with STA/COR1 white-light images to track the CME front, and thus, determine its velocity and acceleration profile. The observing cadence in the EUVI 171~{\AA} filter is 2.5~min for STA and 5~min for STB,
while in the 195~{\AA} filter it is 10~min for both STEREO spacecraft. The white-light STA image cadence is 10~min for COR1 and 15~min for COR2.

The SECCHI\_PREP routine in the SolarSoft software package \citep[SSW;][]{free98} was used to reduce the images.
All images were rescaled to Earth distance to compare observations from STA and STB. STB/EUVI filtergrams were differentially rotated to the same reference time. In addition, the gamma-log function, a compound of the logarithmic and gamma functions, was applied to the STB/EUVI filtergrams to enhance faint structures in the images while retaining detail in the bright features.

\subsection{{\it Hinode} Data}
To derive plasma flow velocities in the active region we use the earliest available EUV raster scan for this event, provided by the EUV Imaging Spectrometer \citep[EIS;][]{culhane07} onboard the \textit{Hinode} spacecraft. The scan covers the period immediately after the decay phase of the CME-associated flare (07:25~--~08:12~UT). The EIS instrument provides high-resolution EUV spectra within two wavebands: 166~--~211~{\AA} and 246~--~292~{\AA}, covering emission lines in the transition region and corona. Four slit/slot positions (1$\arcsec$, 2$\arcsec$, 40$\arcsec$, and 266$\arcsec$) allow a wide choice of operation modes. EIS raster scans are carried out in the following way. In each emission line the intensity is measured by sampling the line across the 1$\arcsec$ or 2$\arcsec$ width of the slit. This sampling is carried out at up to 512~positions across the length of the slit. After this, the slit is moved to the next position of the region of interest until the entire region is scanned. Observations can be carried out in up to 25~lines simultaneously. Images in a particular emission line are constructed as follows. To obtain a pixel column of the image, at each position across the length of the slit, the intensity is integrated over the spectral line width. Repeating this procedure for all slit positions within the region of interest yields the entire image.

After the decay phase of the B2.4 flare, the EIS instrument scanned the active region from west to east in 90 steps, each step 2$\arcsec$ wide, with a slit length of 320$\arcsec$, yielding a field of view (FOV) of $180\arcsec\times320\arcsec$.  The observations comprise eight spectral windows, each 32 pixels wide ({\it i.e.}, the intensity in each line was sampled at 32 positions across the line). We use the Fe~{\sc xii}~195.12~{\AA} window. We calibrated the EIS data using the IDL routine EIS\_PREP, available in the EIS part of the SolarSoft tree. This routine corrects for flat field, dark current, and cosmic rays, and flags saturated, hot, and warm pixels. After this, we derived a Dopplergram from the Fe~{\sc xii}~195.12~{\AA} intensity image as follows. Using the routine EIS\_AUTO\_FIT, we fitted a single Gaussian to the calibrated EIS data to compute a velocity array from the measured line centroid positions. We then corrected the obtained velocity and centroid vectors for the tilt of the EIS slit (EIS\_TILT\_CORRECTION) and the orbital variation of the line centroid (EIS\_ORBIT\_SPLINE).

\subsection{Co-alignment of STEREO/EUVI and \textit{Hinode}/EIS Images}
As aforementioned, EIS images, derived from raster scans, are not snapshots taken at a particular moment in time. When we look at such an image we look at stringed-together pixel columns taken at different instants in time. Therefore, if the scanned region is a flaring region, the features in the FOV may have changed dramatically during the scan. Accordingly, the co-alignment of EIS images and EUVI filtergrams is a tricky issue. Even though there will be similar features in EIS and EUVI images taken in the same passband, the images will not look exactly the same. To solve the problem of co-alignment, we used a method similar to that described by \citet{jin09}. We compared EIS and EUVI images at 195~{\AA}. We set the mid-time of the raster scan as the time of the EIS image, chose an EUVI image observed closest to this time, visually compared it with the EIS image by identifying similar features,
and then shifted the values of XCEN and YCEN in the EUVI image to roughly co-align it with the EIS raster image. After this, we cross-correlated both images to refine the visual co-alignment. We estimate the uncertainty of the co-alignment achieved with this method to $\pm$3$\arcsec$.

\section{Analysis}\label{sec:analysis}

STEREO/EUVI~171 and 195~{\AA} filtergrams reflect the emitting state of the coronal plasma at approx.~1 and 1.5~MK, respectively \citep{wuelser04,howard08}. To investigate the coronal dimming process, we calculate intensity light curves from these filtergrams by (1) summing up at each time all pixel intensities within a region of interest (ROI) and (2) dividing this sum by the number of pixels in this region. For a \textit{global light curve} the ROI is a large rectangular box surrounding the dimming regions, for a \textit{local light curve} the ROI is determined as follows. To enhance the contrast of the dimming regions we use base difference images, {\it i.e.}, we subtract a pre-event image taken around 13-Feb-2009 04:45~UT from all successive frames. We then count among the dim pixels all pixels with intensities in the range
[$I_{\rm{min}}$, $f \cdot I_{\rm{min}}$], where $I_{\rm{min}}$ is always negative, and is
the intensity of the darkest pixel in a difference image and $f \,\epsilon\, [0, 1]$. For instance, for $I_{\rm{min}}=-500$ (in units of normalized detected photons)
and a factor of $f=0.8$, pixels with intensities in the range [$-500$, $-400$] are counted among the dim pixels ({\it i.e.}, an $80$\%-threshold is used). Hence, the number of pixels forming the ROI of a local light curve varies in time. This offers the opportunity to track the size evolution of local dimmings as well. We normalize all light curves to the pre-event intensity level.

The drop in the EUV intensity $I$ during coronal dimmings is usually interpreted as mass loss in the corona in the wake of a CME. For instance, \citet{jin09} showed that the EUV intensity variation in coronal dimming regions reflects the variation of the mass density. Therefore, we consider the intensity change rate $dI/dt$ as a proxy for the mass loss rate, {\it i.e.}, the time evolution of the mass loss. We calculate $dI/dt$ as the time derivative of a spline fit to the normalized dimming light curve.

We derive the temporal evolution of the CME velocity and acceleration ({\it i.e.}, the profiles $v_{\rm{CME}}(t)$ and $a_{\rm{CME}}(t)$, respectively) as follows. We (1) construct running difference images from STA/EUVI 171~{\AA} and COR1 white light frames to enhance the contrast of the CME front, (2) track the fastest-moving part of the front, (3) fit the obtained height-time measurements by a spline fit, and (4) calculate the first and second time derivative of the fit curve.

\section{Results}\label{sec:results}

\subsection{On-Disk EUV Dimmings and Associated CME/Flare Evolution}

\subsubsection{Flare Evolution and Duration of the Dimmings}

\begin{figure*}
 \centerline{\includegraphics[width=0.98\textwidth,clip=]{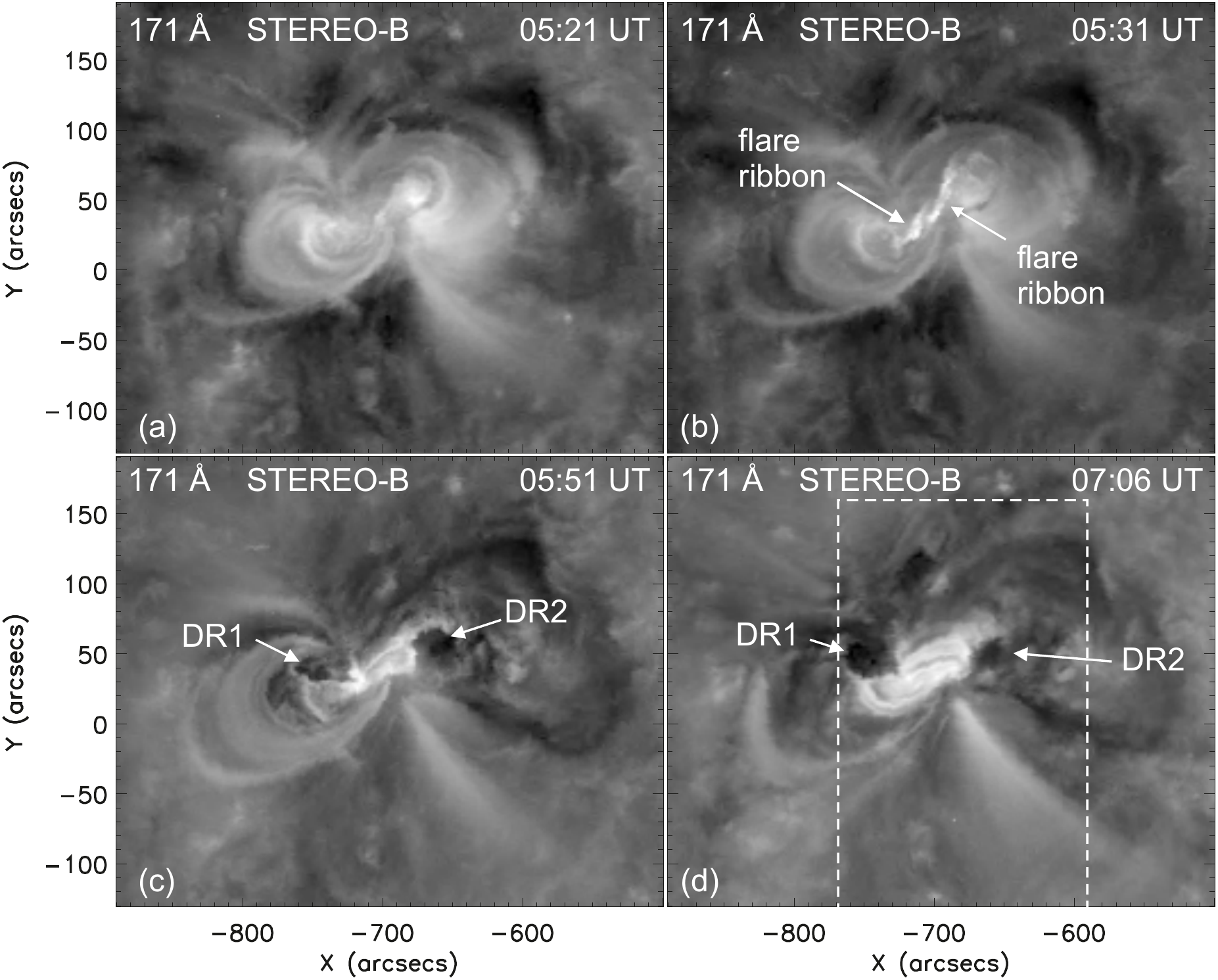}}
 \caption{Evolution of the CME-associated flare and the small-scale coronal dimmings in STB/EUVI~171~{\AA}. (a) Pre-flare S-shaped sigmoid structure. (b) Bright, separating flare ribbons shortly after flare onset. (c) The first postflare loops, resulting from the reconnection of magnetic field lines, appear at the beginning of the decay phase of the flare. The white arrows point to two small-scale dimming regions, designated as DR1 and DR2, which appear near the ends of the pre-flare sigmoid structure. (d) During the decay phase, the dimming regions exhibit an apparent motion. The postflare loops in between the dimmings form an east-west aligned arcade. The dashed box shows the field of view (FOV) of \textit{Hinode}/EIS.~--~North is up, west is to the right.}
    \label{fig:evolution}
\end{figure*}

Figure~\ref{fig:evolution} shows the source region of the CME observed with STB/EUVI~171~{\AA}. The CME-associated flare occurs at the site of an S-shaped sigmoid structure (Figure~\ref{fig:evolution}(a)). During the impulsive phase of the flare, two elongated, roughly north-south aligned flare ribbons separate (Figure~\ref{fig:evolution}(b)). The first postflare loops, resulting from the reconnection of magnetic field lines, appear at the beginning of the decay phase (Figure~\ref{fig:evolution}(c)). Furthermore, two small-scale dimming regions, designated as DR1 and DR2, appear near the ends of the pre-flare sigmoid. These regions grow and exhibit an apparent motion during the decay phase (Figure~\ref{fig:evolution}(d)). The postflare loops in between the dimmings form an east-west aligned arcade.

%  Figure 1

\begin{figure*}
\centerline{\includegraphics[width=0.98\textwidth,clip=]{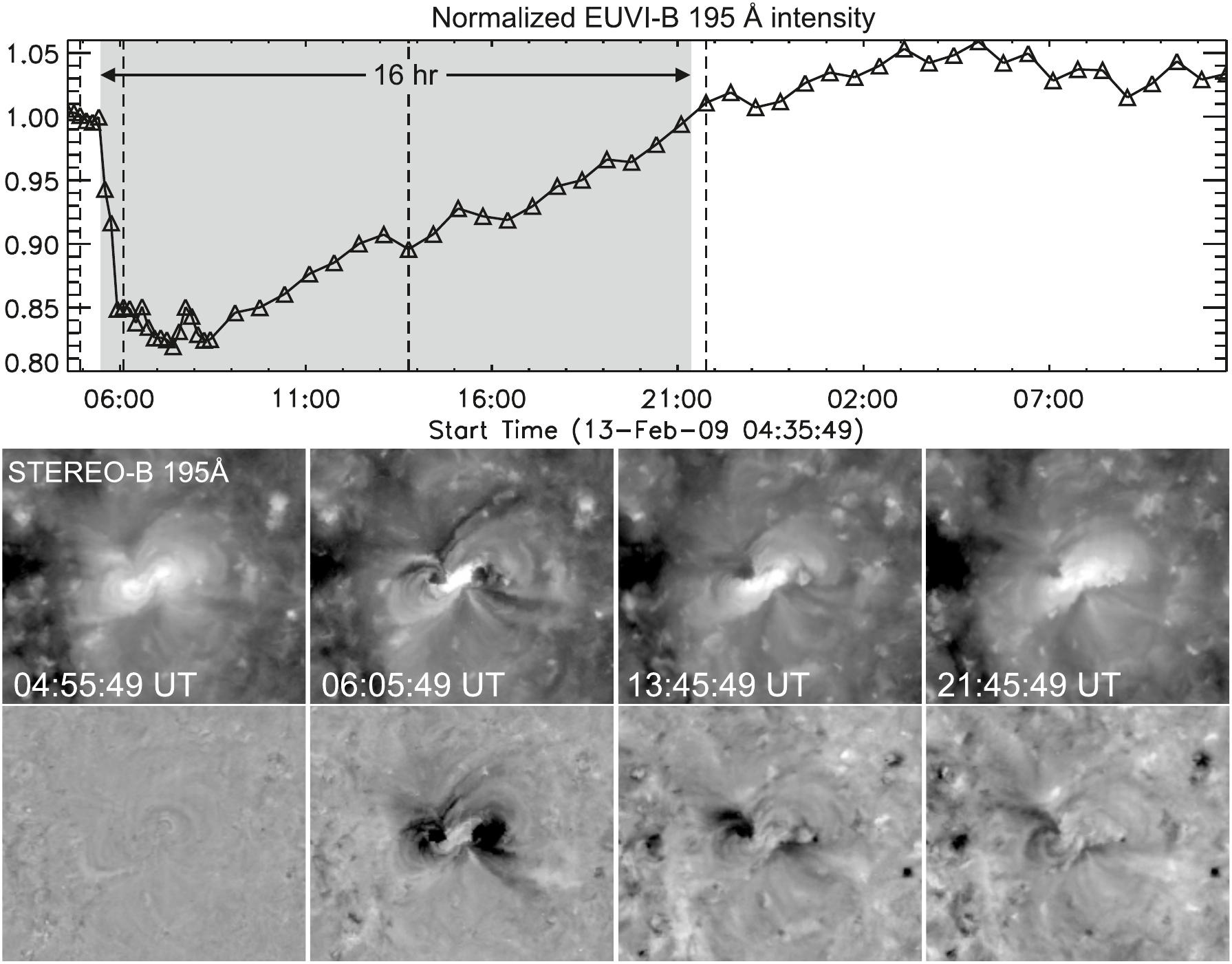}}
\caption{Evolution of the on-disk EUV dimmings. Top: STB/EUVI~195~{\AA} light curve. The intensity values are normalized to the pre-flare EUV flux. The shaded area highlights the period of the dimmings and the EUV flux recovery back to the pre-event level. The times of the images are marked with triangle symbols. The times of the four selected images shown below are marked with dashed lines. Middle: STB/EUVI~195~{\AA} images covering a 600$\arcsec$$\times$500$\arcsec$ FOV centered around the dimmings. Bottom: Corresponding base difference images. At 06:05:49~UT two distinct small-scale dimming regions are visible. They slowly disappear during the next 16~h.}
\label{fig:dimming_duration}
\end{figure*}

In the top row of Figure~\ref{fig:dimming_duration} we plot the STB/EUVI 195~{\AA} intensity light curve calculated from the FOV of the images presented in the middle row (global light curve). At the beginning of the presented time interval we use the full image cadence available (10~min) to capture the abrupt intensity drop. After 13-Feb-2009 08:25~UT the cadence is reduced to 40~min, as a lower cadence is sufficient to capture the subsequent gradual changes in the EUV intensity. The intensity drops down to roughly 83\% of the pre-flare EUV level in approx.~30~min.
The recovery of the EUV intensity back to the pre-event level in the event under study takes approx.~16~h. We note that in extreme cases, recovery
times of coronal dimmings  up to 48~h have been observed  \citep{attrill08}.
The middle and bottom rows show four representative EUVI 195~{\AA} direct and base difference images, respectively, illustrating the onset of the dimmings and the ensuing gradual intensity increase, as the rarefied regions are slowly refilled.

%  Figure 2

\subsubsection{Apparent Motion and Size of the Small-Scale Dimming Regions}\label{subsec:dr1dr2motion}

\begin{figure}
\centerline{\includegraphics[width=0.98\textwidth,clip=]{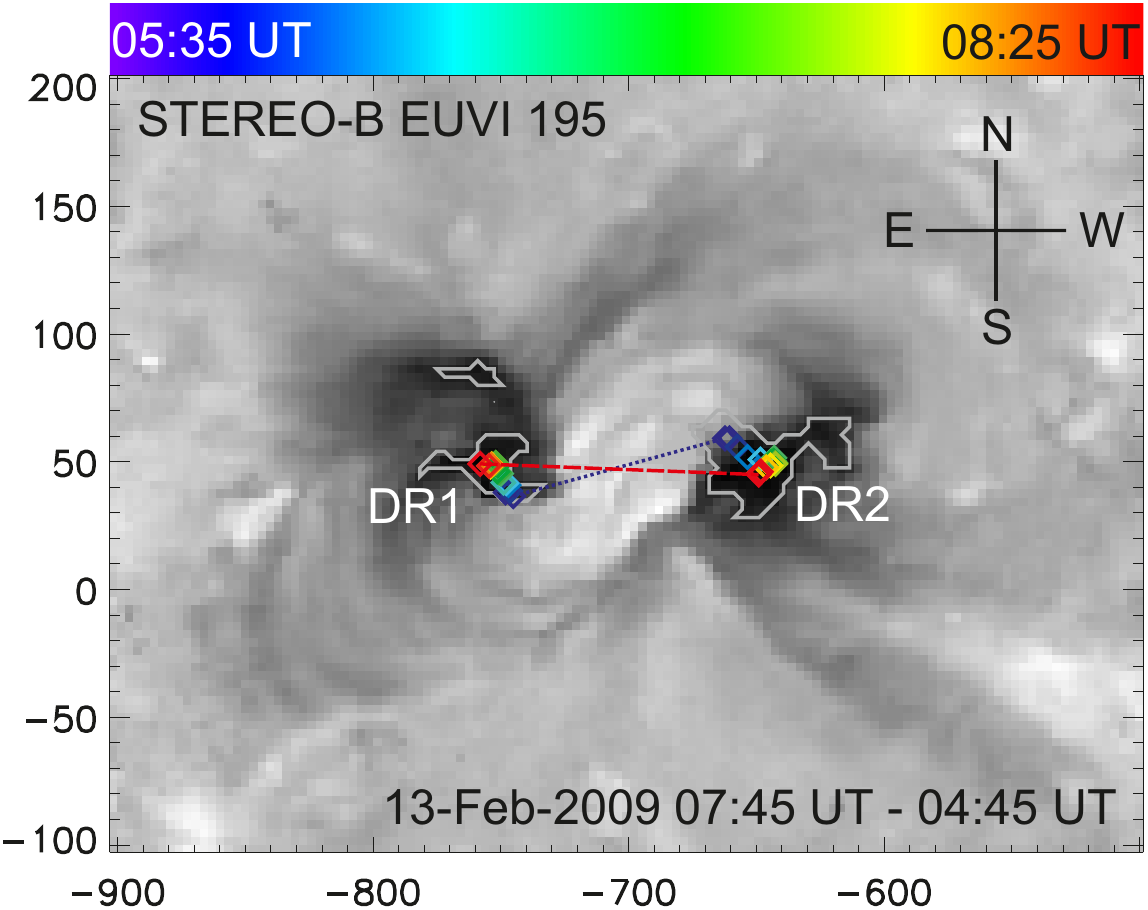}}
\caption[]{STB EUVI 195~{\AA} base difference image of the flaring and localized dimming regions. Bright east-west aligned postflare loops are visible in the center of the image. Gray contours trace the total dimming area, {\it i.e.}, the area that exhibited a drop in intensity within the time interval indicated by the color bar. The contour level represents the 80\%-threshold used to identify the dimming regions. The color-coded diamonds mark the positions of the `center of gravity' (CoG) of dimming regions DR1 and DR2 at each time. The dotted/dashed line connects the CoGs of both dimming regions at the beginning/end of the time interval.}\label{fig:cog_image}
\end{figure}

We use the STB/EUVI~195~{\AA} filtergrams to track the apparent motion of the localized dimming regions DR1 and DR2 ({\it cf.}\ Figure~\ref{fig:evolution}, panels (c) and (d)) and to investigate the evolution of their size. We determine the pixel size of DR1 and DR2 in each base difference image, using a factor of $f = 0.8$. This $80$\%-threshold proved to be most suitable for tracing the dim areas in an image. After this, we calculate the `center of gravity' (CoG) of DR1 and DR2 at each time ({\it i.e.}, for each frame we weight the location of dim pixels by their intensity, and then we divide the sum of these products by the number of dim pixels). Figure~\ref{fig:cog_image} shows the locations of the CoGs for both dimming regions at each time superimposed on a base difference image, taken during the decay phase of the flare. The CoG of DR1 moves to the north-east, while the CoG of DR2 is directed to the south-west. The straight lines connecting the CoGs of both dimming regions at the beginning and end of the analyzed time interval demonstrate that the CoGs rotate clockwise around the `center' of the flare site. Also, taking the east-west alignment of the postflare loops around 07:45~UT as a reference, the lines indicate that the observed rotation reflects the evolution of the dimmings from a `sheared' to a more `potential' configuration, as the event progresses. The successively appearing postflare loops exhibit the same sheared-to-potential evolution ({\it cf.}\ Figure~\ref{fig:evolution}, panels (c) and (d)).

%  Figure 3

\begin{figure}
\centerline{\includegraphics[width=0.98\textwidth,clip=]{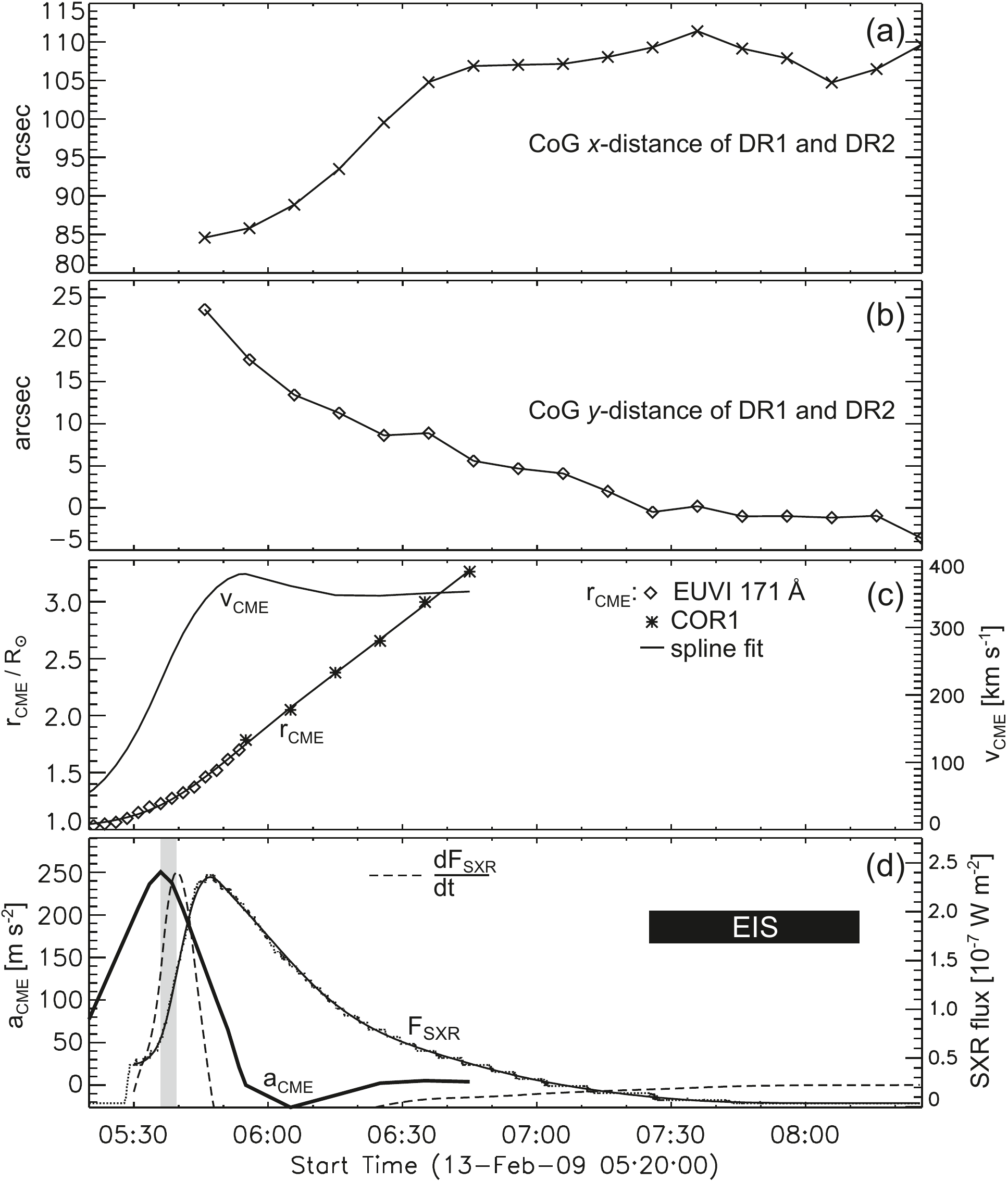}}
\caption[]{Apparent motion of dimming regions DR1 and DR2 in east-west ($x$) and north-south ($y$) direction related to the CME kinematics and the soft X-ray (SXR) emission of the flare. (a) and (b): The $x$- and $y$-distance of the `centers of gravity' (CoG) of DR1 and DR2 observed in STB/EUVI 195~{\AA}. (c): Height-time measurements of the CME front observed in STA/EUVI 171~{\AA} and COR1 white light plus spline fit ($r_{\rm{\rm{CME}}}$) and derived velocity $v_{\rm{\rm{CME}}}$. (d): Derived CME acceleration $a_{\rm{\rm{CME}}}$ (solid thick), GOES12 1~--~8~{\AA} SXR flux (dotted) plus spline fit ($F_{\rm{SXR}}$, solid thin), and time derivative of the fit (dashed). The time derivative is scaled in such a way that its amplitude is equal to the peak of $a_{\rm{\rm{CME}}}$. The vertical bar marks the time difference of 3.5~min between the maxima of $a_{\rm{\rm{CME}}}$ and GOES derivative. The horizontal bar shows the period of the EIS raster scan.}\label{fig:cog}
\end{figure}

In Figure~\ref{fig:cog} we consider the apparent motion of the CoGs of the two dimming regions separately for the east-west ($x$) and north-south ($y$) direction. In addition, we compare the motion of the CoGs with both the kinematics of the CME and the temporal evolution of the soft X-ray emission of the associated flare. The CoGs of the dimming regions separate in the $x$-direction (Figure~\ref{fig:cog}(a)), while they approach each other in the $y$-direction (Figure~\ref{fig:cog}(b)). Given the east-west alignment of the latest appearing postflare loops, the approaching in $y$ reflects the aforementioned evolution of the dimming regions from a `sheared' to a more `potential' configuration. The separation in $x$ indicates that the extent of the region, affected by the expelled CME, grows, as the event progresses. The most conspicuous movement of the CoGs is detected between 05:50~--~06:30~UT ({\it i.e.}, after the impulsive acceleration phase of the CME). In this period, the CME is in its propagation phase, moving with a constant speed of around $350\,\rm{km\,s^{-1}}$ (Figure~\ref{fig:cog}(c)).  After approx.~07:00~UT, the CoGs of both dimming regions seem to stall in place.

In Figure~\ref{fig:cog}(d) we present the CME acceleration profile $a_{\rm{CME}}$. The impulsive acceleration phase of the CME lasts for approx.~40~min and reaches its maximum of $250\,\rm{m\,s^{-2}}$ at 05:36~UT. Around this time, the dimming regions appear and the CoGs start to move. In Figure~\ref{fig:cog}(d) we also investigate the relationship between the acceleration phase of the CME and the energy release phase of the flare. The maximum CME acceleration occurs approx.~11~min before the flare-related maximum of the soft X-ray (SXR) emission. In other words, the highest CME acceleration occurs in the impulsive phase of the flare, during which the flare energy is released and particles are accelerated to nonthermal energies. Since observations of nonthermal emission (hard X-rays, microwaves produced by these particles) are not available for the event under study, we assume that the Neupert effect is valid in this event, {\it i.e.}, we assume that peaks in the time derivative of the SXR flux coincide with peaks in the nonthermal hard X-ray (HXR) emission \citep[{\it e.g.},][]{neup68,dennis93,dennis03,veronig05}. Consequently, we fit a spline curve to the GOES SXR light curve and take the time derivative of the fit to obtain a proxy for the HXR emission, and thus, a proxy for the energy release rate in the flare ({\it cf.}\ dashed profile in Figure~\ref{fig:cog}(d)). The maxima in the SXR time derivative and CME acceleration profile are roughly co-temporal. The time difference between the peaks is less than 4~min. This indicates that the CME acceleration phase is closely synchronized with the energy release phase of the associated flare \citep[{\it e.g.},][]{zhang01,maricic07,vrsnak07,temmer08}.

%  Figure 4

\begin{figure}
\centerline{\includegraphics[width=0.98\textwidth,clip=]{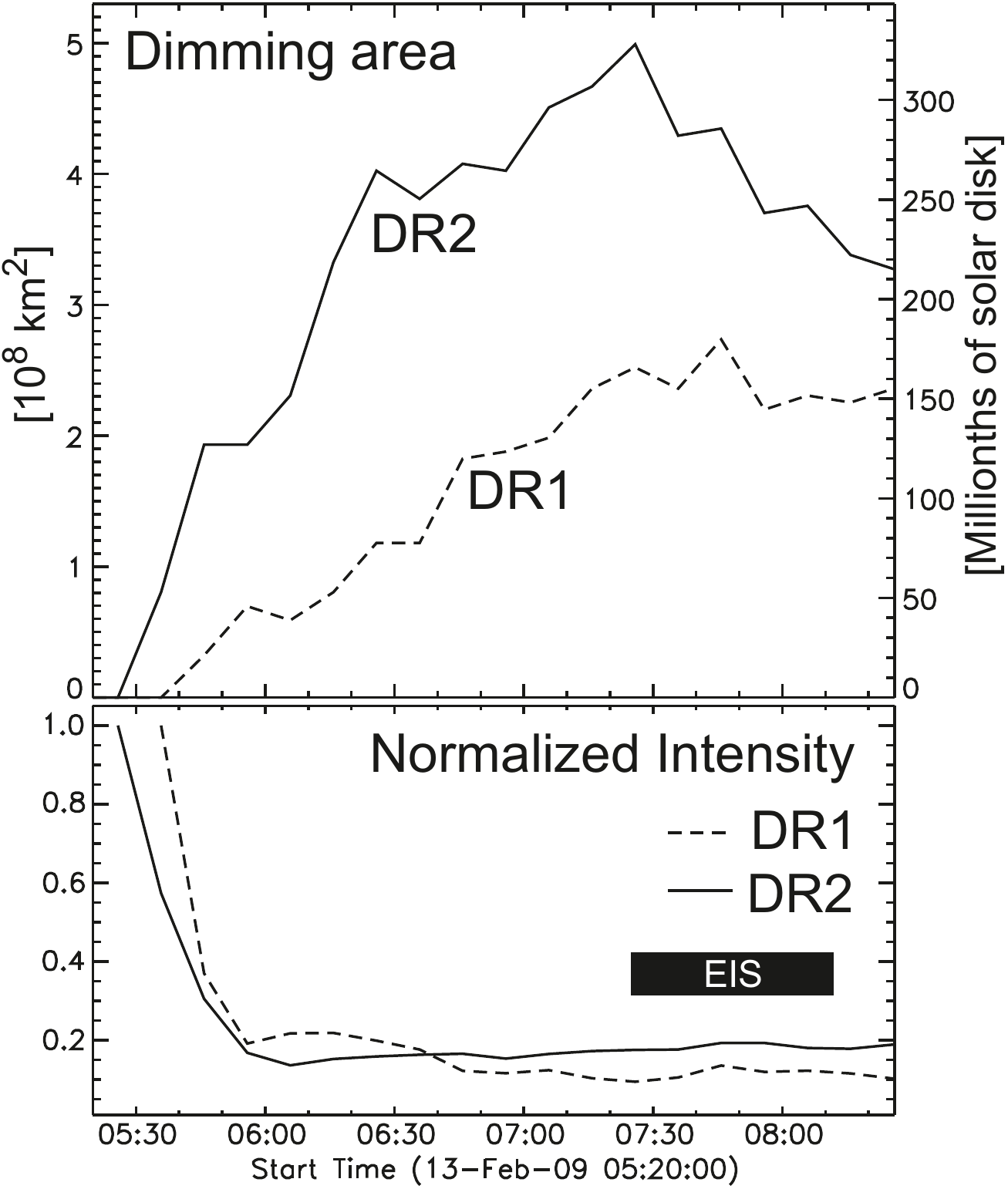}}
\caption[]{Temporal evolution of the dimming areas. Top: Size of dimming areas DR1 and DR2 observed in STB/EUVI 195~{\AA}. Bottom: Normalized intensity, averaged over all pixels in the dimming region. The horizontal bar shows the period of the EIS raster scan.}
\label{fig:DR_size}
\end{figure}

In the upper panel of Figure~\ref{fig:DR_size} we present the temporal evolution of the size of DR1 and DR2. Both dimming regions grow steadily until approx.~07:20~UT, but DR2 grows faster than DR1. After this, the size of DR1 remains almost constant within the time interval under study, while DR2 starts to contract as it is slowly refilled. The bottom panel of Figure~\ref{fig:DR_size} shows the time history of the intensity change of DR1 and DR2, averaged over all pixels in the respective dimming region (local light curves). Both regions undergo an abrupt intensity drop on a timescale of approx.~25--30~min. The intensity drops down to roughly 20\% of the pre-flare EUV flux and remains virtually unchanged during the next 2.5~h.

%  Figure 5

\subsubsection{Dimming Process and Mass Loss Rate Compared to the CME/Flare Progression}\label{subsec:EUV-CME}

\begin{figure*}
\centerline{\includegraphics[width=0.98\textwidth,clip=]{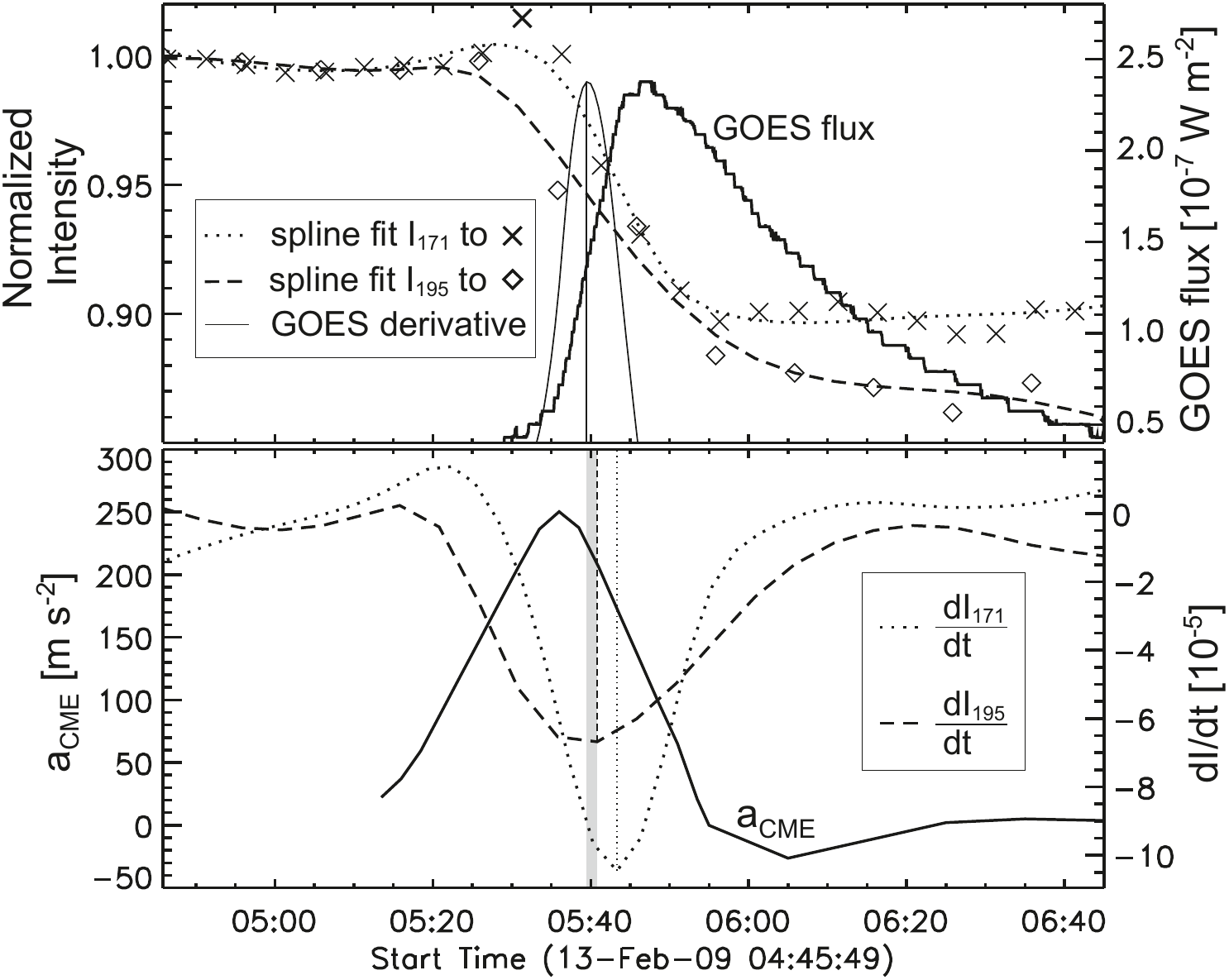}}
\caption[]{Top: GOES12 1~--~8~{\AA} soft X-ray flux (SXR, solid thick), GOES time derivative scaled to the SXR peak flux (solid thin), normalized STB/EUVI 171 and 195~{\AA} light curves integrated over the FOV of Figure~\ref{fig:2methods}b, and spline fits $I_{\rm{171}}$ and $I_{\rm{195}}$  to the data points. The thick $\times$-symbol around 05:32~UT highlights the temporary increase in the 171~{\AA} intensity, caused by the onset of the flare. The vertical line at 05:39:30~UT indicates the maximum of the GOES derivative. Bottom: CME acceleration profile $a_{\rm{CME}}$ (solid) and intensity change rates (time derivatives of the spline curves $I_{\rm{171}}$ and $I_{\rm{195}}$). The vertical lines at 05:40:49 and 05:43:19~UT indicate the times of the highest change in intensity for both filters. The vertical bar marks the time difference of 1.3~min between the maximum of the GOES derivative and the minimum of $dI_{\rm{195}}/dt$.}
\label{fig:dIdt}
\end{figure*}

In the top panel of Figure~\ref{fig:dIdt} we plot: (1) the SXR emission profile of the flare along with its time derivative; (2) the normalized STB/EUVI 171 and 195~{\AA} intensity profiles, calculated over a FOV of 640$\arcsec$$\times$596$\arcsec$ surrounding both the flare site and the small-scale twin dimmings (global light curve); and (3) the spline fits $I$ to the data points. During the impulsive phase of the flare, the intensity drops down to a level of approx.~90\% of the pre-flare EUV flux in 30~min. In the bottom panel of Figure~\ref{fig:dIdt} we investigate the temporal relation between the acceleration phase of the CME (represented by the acceleration profile $a_{\rm{CME}}$) and the EUV intensity change rate $dI/dt$, which is considered as a proxy for the evolution of the mass loss rate ({\it cf.}\ Section~\ref{sec:analysis}). In both wavelengths, the intensity change is most conspicuous during the acceleration phase of the CME. The peak of $a_{\rm CME}$ appears at 05:36~UT. The peaks in the mass loss rate occur a few minutes later, namely, at 05:41~UT in the 195~{\AA} filter and 05:43~UT in the 171~{\AA} filter, respectively. This indicates that large amounts of mass are removed from the analyzed region while the CME is expelled from the Sun. Moreover, the timing of the highest mass loss coincides roughly with the maximum of the SXR time derivative (05:40~UT, {\it cf.}\ Section~\ref{subsec:dr1dr2motion}), suggesting that the energy release in the flare is synchronized with the mass loss due to the CME.

%  Figure 6

\subsection{Plasma Flows in the CME Source Region}\label{sec:plasma_flows}

Large amounts of mass expelled from the Sun in the course of the CME should be reflected in upflowing plasma at the source region of the eruption, in particular, at the locations of the on-disk dimming regions DR1 and DR2, observed with STB/EUVI  195~{\AA} ({\it cf.}\ Figure~\ref{fig:cog_image}). To measure the plasma flow velocities in the dimming regions we use the first available \textit{Hinode}/EIS raster scan and derive a Dopplergram from observations in the Fe~{\sc xii}~195.12~{\AA} emission line.

\begin{figure*}
\centerline{\includegraphics[width=0.98\textwidth,clip=]{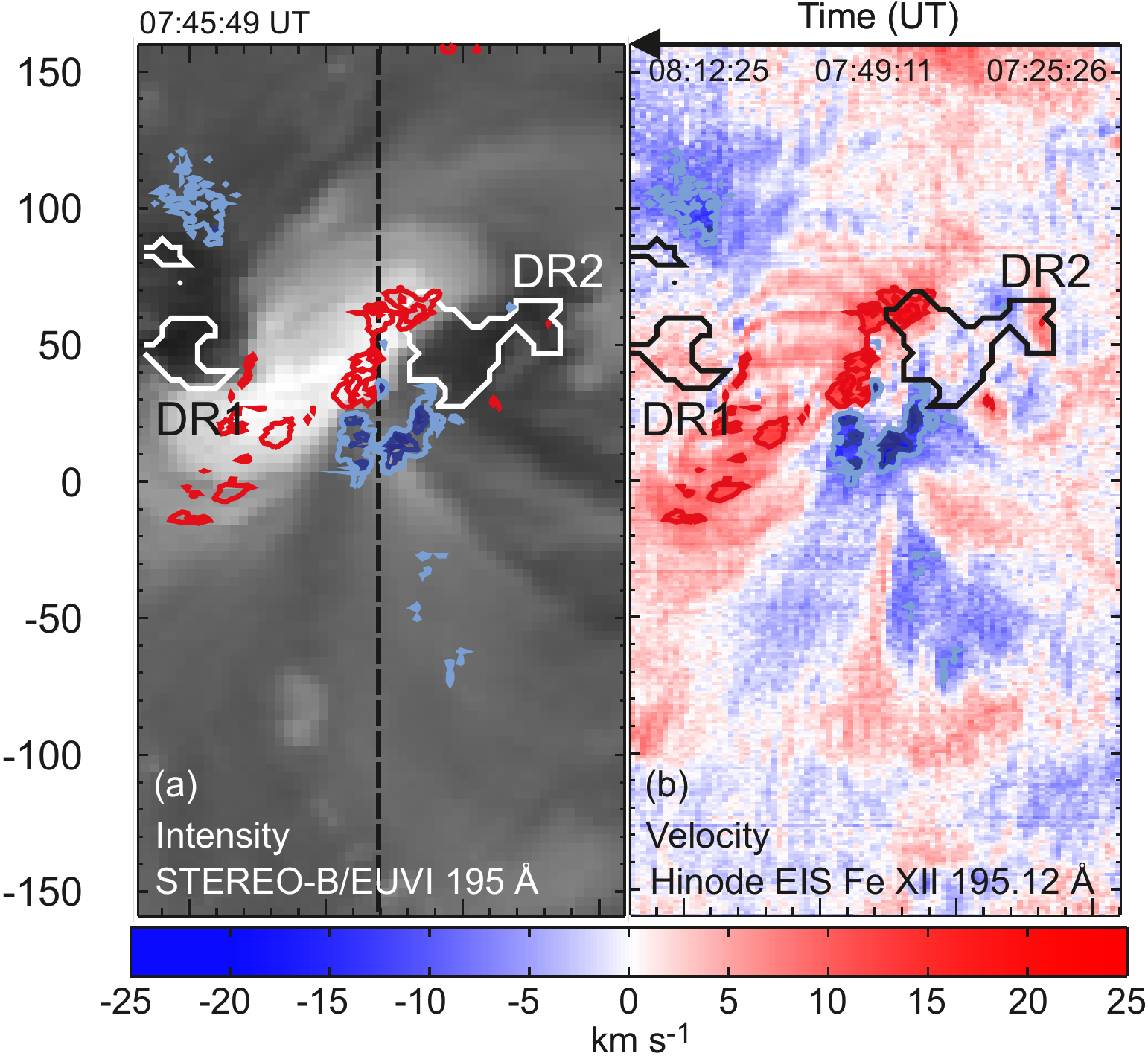}}
\caption[]{(a): STB/EUVI~195~{\AA} intensity image taken at 07:45:49~UT ({\it i.e.}, around the mid-time of the \textit{Hinode}/EIS raster scan), showing the region of the EIS FOV ({\it cf.}\ dashed box in Figure~\ref{fig:evolution}(d)). The dashed line marks the position of the EIS slit at 07:49:11~UT. (b) Dopplergram derived from observations in the \textit{Hinode}/EIS  Fe~{\sc xii}~195.12~{\AA} emission line. Positive and negative velocities correspond to red and blueshifts, respectively. The time axis at the top gives the beginning, mid-time, and end of the EIS raster scan.~--~Both panels: Black/white contours mark the total area of the STB/EUVI~195~{\AA} dimming regions DR1 and DR2 ({\it cf.}\ Figure~\ref{fig:cog_image}). The contour level represents the 80\%-threshold used to identify the dimming regions. The colored contours mark \textit{Hinode}/EIS 195.12~{\AA} plasma flow velocities with contour levels of [$-$25, $-$20, $-$15, $-$10 ]~$\rm{km\,s^{-1}}$ (blue) and [$+10$, $+15$]~$\rm{km\,s^{-1}}$ (red).}
\label{fig:int_vel}
\end{figure*}

Figure~\ref{fig:int_vel}(a) shows an STB/EUVI~195~{\AA} intensity image taken around the mid-time of the EIS raster scan. The FOV is the same as in the EIS instrument ({\it cf.}\ dashed box in Figure~\ref{fig:evolution}(d)). The contours mark the total area of DR1 and DR2 (white) and the EIS flow velocities between $-25$ and $+15\,\rm{km\,s^{-1}}$ (blue/red). The same contours are superimposed on the Dopplergram in Figure~\ref{fig:int_vel}(b). In DR1 we detect only downflows, and these downflows are relatively weak (around $5\,\rm{km\,s^{-1}}$). The strongest downflows with velocities around 15~$\rm{km\,s^{-1}}$ are observed in the arcade of postflare loops, predominantly at the loop footpoints ({\it cf.}\ red contours in Figure~\ref{fig:int_vel}(a)). In DR2 we find both downflows and upflows, but the upflow velocities are low (less than approx.~$5\,\rm{km\,s^{-1}}$).  The strongest upflows with velocities around 20--25~$\rm{km\,s^{-1}}$ are co-spatial with a large-scale, coronal loop structure south-east of DR2 ({\it cf.}\ dark-blue contours in Figure~\ref{fig:int_vel}(a)). This implies that during the EIS raster scan ({\it cf.}\ Figure~\ref{fig:cog}(d)) the strongest upflows of the 1.5~MK plasma seem to be rather flows along large, closed coronal loops than outflows originating from `open' field regions. This is consistent with the finding presented in Section~\ref{subsec:EUV-CME}: The main mass loss, inferred from the STEREO/EUVI intensity drop occurs between roughly 05:25 and 05:55~UT, {\it i.e.}, approx.~90~min before the beginning of the EIS raster scan ({\it cf.}\ Figure~\ref{fig:dIdt} and \ref{fig:DR_size}). During the time interval of the EIS observations, the mass loss rate approaches zero, indicating that the coronal mass loss in the wake of the CME has stopped.

%  Figure 7

\subsection{EUV Dimmings Observed Above the Solar Limb}\label{subsec:EUVI-limb}

\begin{figure*}
\centerline{\includegraphics[width=0.98\textwidth,clip=]{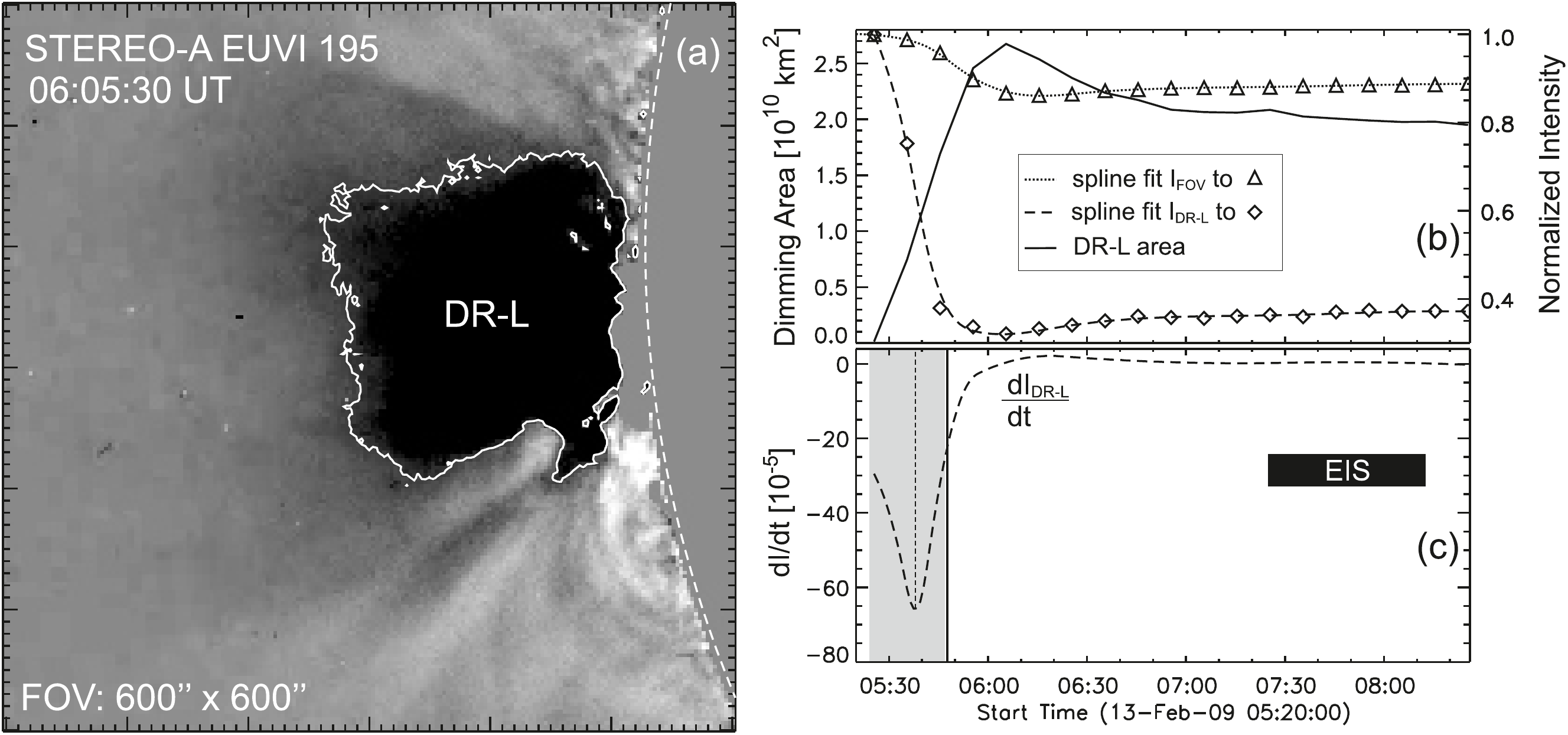}}
\caption[]{EUV dimmings above the limb. (a): STA/EUVI 195~{\AA} difference image with a pre-CME frame around 04:45~UT subtracted from a post-eruption frame at 06:05:30~UT. The solid contour enclosing DR-L represents the 70\%-threshold used to identify the dimming region. The dashed arc marks the solar limb. The solar disk has been artificially occulted. (b) Solid: temporal evolution of the size of DR-L observed in STA/EUVI 195~{\AA}. Dotted: spline fit $I_{\rm{FOV}}$ to the $\Delta$-symbols, which represent the normalized intensity calculated from the FOV of panel (a) at each time. Dashed: spline fit $I_{\rm{DR-L}}$ to the $\diamond$-symbols, which represent the normalized intensity calculated from all pixels forming DR-L at each time. (c): Time derivative of the spline fit $I_{\rm{DR-L}}$. The gray vertical bar highlights the period of strongest CME acceleration, represented by the FWHM of the CME acceleration profile shown in Figure~\ref{fig:cog}(d). The vertical lines mark the time of the minimum of the intensity change rate at 05:37:30~UT (dashed) and the time of the GOES maximum of the flare at 05:47:39~UT (solid). The horizontal black bar highlights the period of the EIS raster scan.}
\label{fig:limb_dimmings}
\end{figure*}

We use the STA/EUVI~195~{\AA} filtergrams to investigate the dimming process above the solar limb. Figure~\ref{fig:limb_dimmings}(a) shows the extent of the dimming region, designated as DR-L, at 06:05~UT ({\it i.e.}, in the early decay phase of the flare). A 70\%-threshold was used to determine the pixel size of DR-L at each time ($f = 0.7$). In Figure~\ref{fig:limb_dimmings}(b) we present both a local and a global normalized intensity light curve. The global profile is calculated from the entire FOV of Figure~\ref{fig:limb_dimmings}(a), while the local one is derived from all pixels forming DR-L at each time. The spline fits to the profiles show that the EUV intensity decrease is more distinct inside DR-L than in the entire FOV (intensity drop to 32\% vs.~86\% of the pre-CME flux in 25 vs.~40~min). After 06:00~UT, a slight gradual intensity increase is detected in both profiles. Figure~\ref{fig:limb_dimmings}(b) also shows the size evolution of DR-L. The dim area grows steadily during the period of intensity drop and reaches its maximum size around 06:05~UT. After this, the area contracts slowly for about an hour, and then its size remains virtually unchanged within the time interval under study ({\it i.e.}, until 08:25~UT).

Figure~\ref{fig:limb_dimmings}(c) shows the time derivative of the local intensity profile $I_{\rm{DR-L}}$. The curve progression indicates that the intensity drop, and therefore mass loss, is most distinct during the period highlighted by the shaded box in Figure~\ref{fig:limb_dimmings}(c). The width of this box is the FWHM of the CME acceleration profile shown in Figure~\ref{fig:cog}(d) and in the bottom panel of Figure~\ref{fig:dIdt}. We use the FWHM to represent the period of strongest CME acceleration. The end of this period roughly coincides with the end of the impulsive phase of the flare ({\it cf.}\ vertical solid line at around 05:48~UT in Figure~\ref{fig:limb_dimmings}(c), marking the GOES maximum). The peak time of the intensity change rate at 05:37~UT is close to the middle of the FWHM box (05:36~UT). After the period of strongest CME acceleration the intensity change rate approaches zero, in agreement with the intensity change rate derived from the on-disk observations ({\it cf.}\ Section~\ref{subsec:EUV-CME}). This indicates that the mass loss is synchronized with the CME acceleration phase and substantial only during the period of strongest CME acceleration.

%  Figure 8

\subsection{CME Velocity in the Initial Stage of the Eruption}\label{subsubsec:velocity}

%  Figure 9
\begin{figure*}
\centerline{\includegraphics[width=0.98\textwidth,clip=]{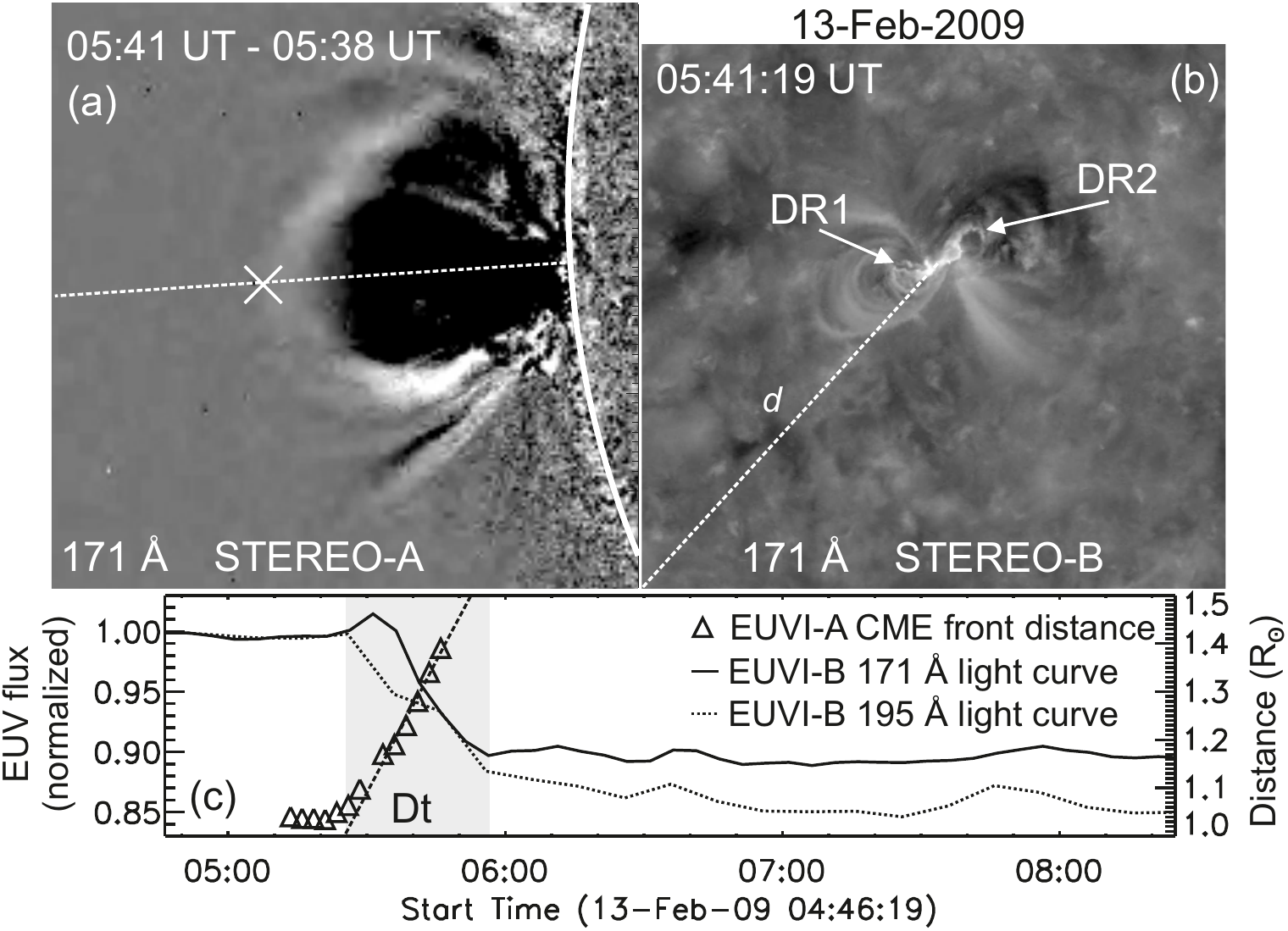}}
\caption[]{(a): STA/EUVI~171~{\AA} running difference image. Dashed line: direction, in which the CME front was tracked; $\times$-symbol: one of the height-time measurements; solid arc: solar limb. (b): STB/EUVI~171~{\AA} filtergram taken at the same time as (a). The dashed line from the flare site to the most remote vertex of the FOV marks the distance $d$, used to estimate the average CME velocity. Arrows point to dimming regions DR1 and DR2. FOV: 640$\arcsec$$\times$596$\arcsec$. (c): Solid/dotted: STB/EUVI 171/195~{\AA} flux integrated over the FOV of panel (b) and normalized to the pre-flare intensity. The $\Delta$-symbols represent the averaged STA height-time measurements of the CME front. The dashed line is the linear fit to the data points inside the gray vertical bar that highlights the period $\Delta t$ of the EUV intensity drop.}\label{fig:2methods}
\end{figure*}

In Figure~\ref{fig:cog}(c) we presented the evolution of the CME velocity, derived from combined STA/EUVI and COR1 above-the-limb observations, extending over the period between roughly 05:20 and 06:45~UT. In the following, we exploit the availability of both above-the-limb and on-disk EUVI observations in one and the same event to compare two methods of determining the velocity of the CME in the initial stage of the eruption ({\it i.e.}, until approx.~05:56~UT).

In the first method we use \textit{above-the-limb} observations. We apply the procedure described in Section~\ref{sec:analysis}, {\it i.e.}, we track the fastest part of the CME front in STA/EUVI 171~{\AA} running difference images and calculate the velocity from a linear fit to the height-time measurements. Figure~\ref{fig:2methods}(a) shows an example of such a measurement. To estimate the error we repeat the measuring procedure several times and determine the mean and standard deviation of the calculated velocities. Thus, we obtain a linear CME velocity of $212 \pm 10\,\rm{km}\,s^{-1}$ between 05:26 and 05:46~UT.

In the second method we use STB/EUVI 171 and 195~{\AA} \textit{on-disk} observations of the region surrounding both the flare site and the small-scale twin dimmings. To select an appropriate FOV, {\it i.e.}, a FOV covering the range over which the outermost border of the expanding global-scale dimmings is discernible, we visually track these dimmings in difference images. Figure~\ref{fig:2methods}(b) shows the FOV selected according to this criterion. We associate the expanding global-scale dimmings with the 1~MK and 1.5~MK plasma, rarefied or ejected in the course of the CME \citep[{\it cf.},][]{aschwanden99}. We calculate the intensity profiles for both wavelengths, integrating over the entire FOV of Figure~\ref{fig:2methods}(b) (global light curve). We then normalize both profiles to the pre-flare intensity. Figure~\ref{fig:2methods}(c) shows that the intensity drops down to a level of 87--90\% of the pre-flare EUV flux on a timescale of $\Delta t \approx 30$~min (05:26~--~05:56~UT). This is the period over which the CME clears the FOV, because afterwards the intensity level in both wavelengths stabilizes. The average velocity of the CME in this period is calculated as $v_{\rm{CME}} = d/\Delta t$, where $d$ ($\approx 345,000\,\rm{km}$) is the distance shown in Figure~\ref{fig:2methods}(b). To estimate $\Delta t$ we use both the EUVI 171~and 195~{\AA} light curves and determine the times when the intensity drop in both profiles starts and ends ({\it cf.}\ shaded box in Figure~\ref{fig:2methods}(c), highlighting the time interval $\Delta t$). This yields an average speed of $v_{\rm{CME}}  \approx 192 \,\rm{km\,s^{-1}}$. Due to the image cadence of 5~min for STB 171~{\AA} and 10~min for STB 195~{\AA}, however, $\Delta t$ is likely to be overestimated, since the actual intensity drop may start later, namely, between the time of the last pre-event intensity image and the time of the first `intensity-drop' image. Similarly, the actual intensity drop may end earlier, namely, between the time of the last 'intensity-drop' image and the time of the first image with stabilized intensity. Assuming that the overestimation of $\Delta t$ is up to 10~min, we obtain an average speed of $v_{\rm{CME}}  \approx 265 \,\rm{km\,s^{-1}}$. Taking the mean of both velocity values we obtain an average speed of $v_{\rm{CME}} \approx 228 \pm 36\,\rm{km\,s^{-1}}$.

Both the `above-the-limb' method (CME tracking) and the `on-disk' method (expanding global-scale dimming) yield comparable results.

\section{Discussion}\label{sec:discussion}

Taking advantage of the quadrature of the STEREO Ahead and Behind spacecraft on 13 February 2009, we presented the first direct comparison of CME-associated, coronal EUV dimmings observed simultaneously on the solar disk and above the solar limb. Both data sets yield a consistent picture of the eruption and mass loss during the early phase of the CME, and thus, support the view that above-the-limb dimmings and global-scale, on-disk dimmings essentially reflect the same process, namely, the density rarefication in the corona due to the expelled CME.

We compared the dimming process with the kinematics of the CME and found that during the period of strongest CME acceleration (approx.~05:25--05:50~UT) the EUV intensity in both data sets dropped abruptly. Also, the mass loss rates, inferred from both on-disk and above-the-limb EUVI intensity observations, showed a distinct extremum in this period, suggesting that the bulk of the coronal mass was ejected, as the CME left the Sun. Moreover, the extrema of the CME acceleration profile and mass loss rates were only a few minutes apart in time, reflecting the close temporal relationship between the kinematics of the CME and the dimming process. Two hours after the eruption, both the EUVI~195~{\AA} mass loss rates and the Dopplergram derived from Fe~{\sc xii}~195.12~{\AA} observations with the \textit{Hinode}/EIS instrument indicated that the loss of coronal plasma with a temperature of 1.5~MK had virtually stopped.

The main characteristics of the dimmings in the event under study were comparable with other findings. For example, an intensity decrease within 20--30~min and a gradual recovery of the EUV brightness lasting several hours are characteristic features of the coronal dimming process \citep[{\it e.g.},][]{aschwanden99,mcintosh07,reinard08,aschwanden09a}.
Further, the observed amount of intensity drop, including the difference in brightness decrease between small-scale and global-scale dimmings, is typical for the two types of dimmings \citep{aschwanden99,zarro99,thomson00,mcintosh07,harra07}. Also, the initial growing of dimming areas as well as the apparent motion of these regions away from the source region of the eruption was reported earlier \citep{mcintosh07}. In the 13 February 2009 CME event, however, we observed for the first time an additional motion pattern of the small-scale twin dimming regions, namely, an apparent clockwise rotation around the source region of the eruption or the `center' of the flare site, respectively. Taking the orientation of the first and last appearing postflare loops as a reference, this rotation can also be described as a 'sheared-to-potential' migration, superposed on the outward motion.

How can we explain such an apparent rotation? Small-scale twin dimmings are often interpreted as the confined footpoint areas of a CME-associated, erupting flux rope \citep{sterling97,webb00}. The footpoints of a flux rope, however, are stationary. Therefore, moving bipolar dimmings represent not only the footpoints of the flux rope, but also other regions, where field lines are stretched. From MHD simulations the destabilization and subsequent eruption of flux ropes/CMEs is often achieved by twisting magnetic field structures \citep[see {\it e.g.}][]{chen02,fan04,torok03}. Associated dimming regions can as well be located at large distances from the source region of the eruption. For example, \citet{delannee00} reported dimmings at the footpoints of transequatorial loops that connected a flaring active region to an active region located in the opposite hemisphere. They proposed that the appearance of such dimmings is strongly related to the magnetic field topology: When low-lying, active-region loops rise and erupt, thus forming a characteristic twin dimming pattern, the overlying large-scale loops rise and erupt as well. \citet {mandrini07} described a scenario in which the arcade above the erupting flux rope expands significantly before reconnecting. In this case, twin dimmings are expected to appear not only at the footpoints of the erupting flux rope, but also at the footpoints of the sheared magnetic arcade, as the expansion occurs.

\begin{figure*}
\centerline{\includegraphics[width=0.98\textwidth,clip=]{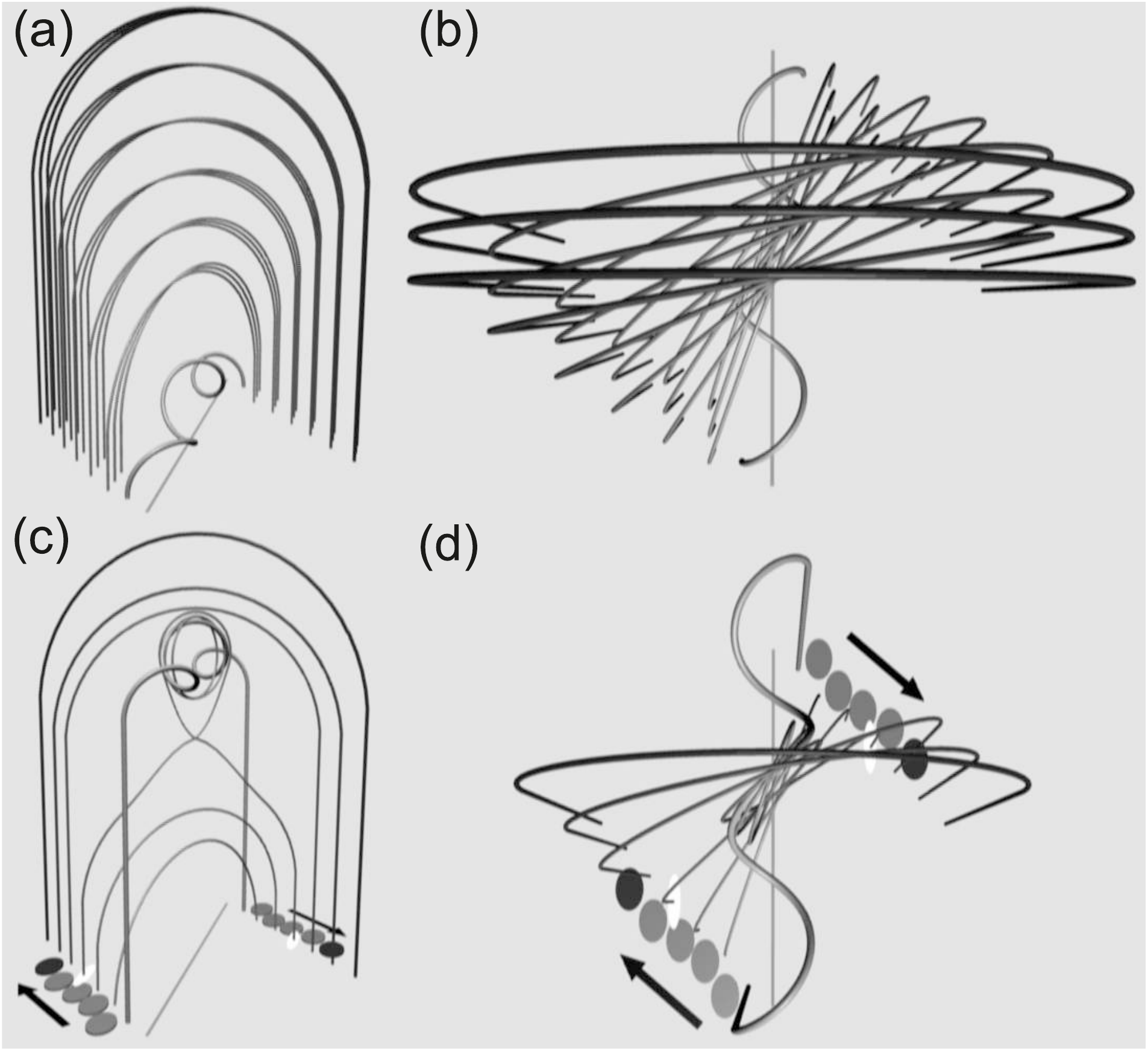}}
\caption{Apparent migration of the CoGs of the localized dimming regions. Top row: side view (a) and top view (b) before the eruption. The flux rope (helical structure) is rooted at both sides of the PIL (straight line). Close to the PIL the overlying arcade consists of low-lying, strongly sheared loops, whereas at greater distances from the PIL the amount of shearing decreases and the altitude of the loop tops increases. Bottom row: side view (c) and top view (d) during the eruption. For clarity, only one field line of each field line triple from (a) and (b) is shown. The field lines of the overlying arcade are pushed upward by the erupting flux rope. Shortly before the thus stretched field lines reconnect, dim patches appear near their footpoints. The dark, filled ellipses mark the positions of the CoGs of the coronal dimmings at different times (black/gray: CoGs near currently/recently stretched lines). The arrows point in the direction of the apparent motion of the CoGs. The white ellipses highlight the bright flare ribbons (energy deposition sites) at the footpoints of the currently reconnected field lines.}\label{fig:DRMovement}
\end{figure*}

Bearing in mind this idea we propose a magnetic field topology in the event under study that may explain the observed rotation and outward migration of the small-scale twin dimming regions DR1 and DR2 as well as the appearance of less and less sheared postflare loops, as the eruption progresses ({\it cf.}\ Figure~\ref{fig:DRMovement}). Before the eruption, a magnetic separatrix layer divides the flux rope from the overlying arcade that spans the polarity inversion line (PIL). The inner loops of this arcade, {\it i.e.}, loops that are rooted closer to the PIL, are strongly sheared, low-lying loops, whereas the loop tops of the less and less sheared outer loops are at successively higher altitudes. Once the magnetic configuration has become unstable the flux rope erupts. As a consequence, the field lines of the flux rope are stretched and the first dimmings appear near its footpoints, {\it i.e.}, at strongly sheared positions. When the erupting flux rope encounters the field lines of the arcade, it stretches them while it continues to rise. Shortly before the stretched field lines reconnect, dimmings appear near their footpoints. Since the loop tops of the overlying arcade are located at different heights, the low-lying, inner loops, rooted at strongly sheared positions, are stretched before the large-scale, outer loops, whose footpoints are less and less sheared. Therefore, the respective dimmings appear at different times and positions, and the observer gets the impression of rotating twin dimming regions or dimmings that evolve from a `sheared' to a more `potential' configuration while they move away from the source region of the eruption, respectively. Once the stretched field lines of the overlying arcade reconnect, their upper parts supply the erupting flux rope with new poloidal flux, while their lower parts form the postflare loops. Since the first appearing postflare loops are created from the low-lying, inner, and strongly sheared loops, they are strongly sheared themselves, whereas the later appearing postflare loops are created from the less and less sheared outer loops, and thus, have a more and more potential configuration.

%  Figure 10

In analyzing the behavior of the relatively small twin dimmings DR1 and DR2 on the solar surface we reach the conclusion that the observed motion pattern reflects not only the eruption of the flux rope, but also the ensuing stretching of the overlying arcade. The above-the-limb dimming region DR-L, however, is larger than DR1 and DR2 by a factor of 100 ({\it cf.}\ top panel of Figure~\ref{fig:DR_size} and Figure~\ref{fig:limb_dimmings}(b)), and thus, rather represents the vast cross section of the coronal volume, evacuated by the CME.

Finally, the existence of both STA and STB/EUVI observations of the 13 February 2009 CME event allowed us for the first time to compare directly two methods of deriving the CME velocity in the initial stage of the eruption. The velocity values obtained from above-the-limb observations of the leading edge of the CME (CME tracking) and on-disk observations of the expanding global-scale dimmings were comparable ($212 \pm 10\,\rm{km}\,s^{-1}$ and $228 \pm 36\,\rm{km\,s^{-1}}$, respectively). This demonstrates that global-scale, on-disk dimmings, originating from the source region of the eruption, propagate with a speed similar to that of the leaving CME front, and thus, supports the general view that global-scale EUV dimmings are on-disk signatures of CME launches. Accordingly, they can be used as a means for identifying CME source regions and estimating initial CME velocities, which is of particular interest in connection with Earth-directed halo CMEs and space weather forecast.

\begin{acks}
This study was supported by the Austrian Austrian Science Fund (FWF): P20145-N16. {\it Hinode} is a Japanese mission developed and launched by ISAS/JAXA, with NAOJ as domestic partner and NASA and STFC (UK) as international partners. It is operated by these agencies in co-operation with ESA and NSC (Norway). The {\rm SECCHI} data used here were produced by an international consortium of the {\it Naval Research Laboratory} (USA), {\it Lockheed Martin Solar and Astrophysics Lab} (USA), {\it NASA Goddard Space Flight Center} (USA), {\it Rutherford Appleton Laboratory} (UK), {\it University of Birmingham} (UK), {\it Max-Planck-Institut for Solar System Research} (Germany), {\it Centre Spatiale de Liege} (Belgium), {\it Institut d'Optique Theorique et Appliquee} (France), and {\it Institut d'Astrophysique Spatiale} (France).
\end{acks}

%%% %%%%%%%%%%%%%%%%%%%%%%%%%%%%%%%%%%%%%%%%%%%%%%%%%%%%%%%%%%%
%% Bibliography
%
% Using BibTeX
%
%

\bibliographystyle{spr-mp-sola-cnd}

\begin{thebibliography}{50}
\expandafter\ifx\csname natexlab\endcsname\relax\def\natexlab#1{#1}\fi

\bibitem[{{Aschwanden} {\it et al.}(1999){Aschwanden}, {Fletcher}, {Schrijver}, \&
  {Alexander}}]{aschwanden99}
{Aschwanden}, M.~J., {Fletcher}, L., {Schrijver}, C.~J., {Alexander}, D.:
  1999, {\apj} {\bf 520}, 880.

\bibitem[{{Aschwanden} {\it et al.}(2009{\natexlab{a}}){Aschwanden}, {Nitta},
  {Wuelser}, {Lemen}, {Sandman}, {Vourlidas}, \& {Colaninno}}]{aschwanden09b}
{Aschwanden}, M.~J., {Nitta}, N.~V., {Wuelser}, J., {Lemen}, J.~R., {Sandman},
  A., {Vourlidas}, A., {Colaninno}, R.~C.: 2009{\natexlab{a}},
{\apj} {\bf 706}, 376.

\bibitem[{{Aschwanden} {\it et al.}(2009{\natexlab{b}}){Aschwanden}, {Wuelser},
  {Nitta}, \& {Lemen}}]{aschwanden09a}
{Aschwanden}, M.~J., {Wuelser}, J.~P., {Nitta}, N.~V., {Lemen}, J.~R.:
  2009{\natexlab{b}}, {\solphys} {\bf 256}, 3.

\bibitem[{{Attrill} {\it et al.}(2008)}]{attrill08}
{Attrill}, G.~D.~R., {van Driel-Gesztelyi}, L., {D{\'e}moulin}, P.,
{Zhukov}, A.~N., {Steed}, K., {Harra}, L.~K., {Mandrini}, C.~H.,
{Linker}, J.: 2008, {\solphys} {\bf 252}, 349.


\bibitem[{{Bewsher} {\it et al.}(2008){Bewsher}, {Harrison}, \&
  {Brown}}]{bewsher08}
{Bewsher}, D., {Harrison}, R.~A., {Brown}, D.~S.: 2008, {\aap} {\bf 478}, 897.


\bibitem[{{Cohen} {\it et al.}(2009){Cohen}, {Attrill}, {Manchester}, \&
  {Wills-Davey}}]{cohen09}
{Cohen}, O., {Attrill}, G.~D.~R., {Manchester}, W.~B., {Wills-Davey}, M.~J.:
  2009, {\apj} {\bf 705}, 587.


\bibitem[{{Chen}(1989)}]{chen89}
{Chen}, J.: 1989, {\apj} {\bf 338}, 453.

\bibitem[{{Chen} {\it et al.}(2002){Chen}, {Wu}, {Shibata}, \& {Fang}}]{chen02}
{Chen}, P.-F., Wu, S.~T., Shibata, K., Fang, C.: 2002, {\apjl} {\bf 572}, 99.



\bibitem[{{Culhane} {\it et al.}(2007){Culhane}, {Harra}, {James}, {Al-Janabi},
  {Bradley}, {Chaudry}, {Rees}, {Tandy}, {Thomas}, {Whillock}, {Winter},
  {Doschek}, {Korendyke}, {Brown}, {Myers}, {Mariska}, {Seely}, {Lang}, {Kent},
  {Shaughnessy}, {Young}, {Simnett}, {Castelli}, {Mahmoud}, {Mapson-Menard},
  {Probyn}, {Thomas}, {Davila}, {Dere}, {Windt}, {Shea}, {Hagood}, {Moye},
  {Hara}, {Watanabe}, {Matsuzaki}, {Kosugi}, {Hansteen}, \&
  {Wikstol}}]{culhane07}
{Culhane}, J.~L., Harra, L.~K., James, A.~M., Al-Janabi, K.,
Bradley, L.~J., Chaudry, R.~A., {\it et~al.}: 2007, {\solphys} {\bf 243}, 19.


\bibitem[{{Delann{\'e}e}(2000)}]{delannee00}
{Delann{\'e}e}, C.: 2000, {\apj} {\bf  545}, 512.

\bibitem[{{Dennis} {\it et al.}(2003){Dennis}, {Veronig}, {Schwartz}, {Sui},
  {Tolbert}, {Zarro}, \& {Rhessi Team}}]{dennis03}
{Dennis}, B.~R., {Veronig}, A., {Schwartz}, R.~A., {Sui}, L., {Tolbert}, A.~K.,
  {Zarro}, D.~M., {Rhessi Team}: 2003, {\it Adv. Space Res.} {\bf 32}, 2459.

\bibitem[{{Dennis} and {Zarro}(1993)}]{dennis93}
{Dennis}, B.~R., {Zarro}, D.~M.: 1993, {\solphys} {\bf 146}, 177.

\bibitem[{{Eyles} {\it et al.}(2009){Eyles}, {Harrison}, {Davis}, {Waltham},
  {Shaughnessy}, {Mapson-Menard}, {Bewsher}, {Crothers}, {Davies}, {Simnett},
  {Howard}, {Moses}, {Newmark}, {Socker}, {Halain}, {Defise}, {Mazy}, \&
  {Rochus}}]{eyl09}
{Eyles}, C.~J., Harrison, R.~A., Davis, C.~J., Waltham, N.~R.,
Shaughnessy, B.~M., Mapson-Menard, H.~C.~A., {\it et~al.}: 2009,
{\solphys} {\bf 254}, 387.


\bibitem[{{Fan} and {Gibson}(2004)}]{fan04}
{Fan}, Y., {Gibson}, S.~E.: 2004, {\apj} {\bf 609}, 1123.

\bibitem[{{Freeland} and {Handy}(1998)}]{free98}
{Freeland}, S.~L., {Handy}, B.~N.: 1998, {\solphys} {\bf 182}, 497.

\bibitem[{{Galvin} {\it et al.}(2008){Galvin}, {Kistler}, {Popecki}, {Farrugia},
  {Simunac}, {Ellis}, {M{\"o}bius}, {Lee}, {Boehm}, {Carroll}, {Crawshaw},
  {Conti}, {Demaine}, {Ellis}, {Gaidos}, {Googins}, {Granoff}, {Gustafson},
  {Heirtzler}, {King}, {Knauss}, {Levasseur}, {Longworth}, {Singer}, {Turco},
  {Vachon}, {Vosbury}, {Widholm}, {Blush}, {Karrer}, {Bochsler}, {Daoudi},
  {Etter}, {Fischer}, {Jost}, {Opitz}, {Sigrist}, {Wurz}, {Klecker}, {Ertl},
  {Seidenschwang}, {Wimmer-Schweingruber}, {Koeten}, {Thompson}, \&
  {Steinfeld}}]{gal08}
{Galvin}, A.~B., Kistler, L.~M., Popecki, M.~A., Farrugia, C.~J.,
Simunac, K.~D.~C., Ellis, L., {\it et~al.}: 2008,
{\it Space Sci. Rev.} {\bf 136}, 437.

\bibitem[{{Harra} {\it et al.}(2007){Harra}, {Hara}, {Imada}, {Young}, {Williams},
  {Sterling}, {Korendyke}, \& {Attrill}}]{harra07}
{Harra}, L.~K., {Hara}, H., {Imada}, S., {Young}, P.~R., {Williams}, D.~R.,
  {Sterling}, A.~C., {Korendyke}, C., {Attrill}, G.~D.~R.: 2007, {\pasj} {\bf 59}, 801.

\bibitem[{{Harra} and {Sterling}(2001)}]{harra01}
{Harra}, L.~K., {Sterling}, A.~C.: 2001, {\apjl} {\bf 561}, 215.


\bibitem[{{Harrison} {\it et al.}(2003){Harrison}, {Bryans}, {Simnett}, \&
  {Lyons}}]{harrison03}
{Harrison}, R.~A., {Bryans}, P., {Simnett}, G.~M., {Lyons}, M.: 2003,
{\aap} {\bf 400}, 1071.

\bibitem[{{Harrison} and {Lyons}(2000)}]{harrison00}
{Harrison}, R.~A., {Lyons}, M.: 2000, {\aap} {\bf 358}, 1097.

\bibitem[{{Howard} {\it et al.}(2008){Howard}, {Moses}, {Vourlidas}, {Newmark},
  {Socker}, {Plunkett}, {Korendyke}, {Cook}, {Hurley}, {Davila}, {Thompson},
  {St Cyr}, {Mentzell}, {Mehalick}, {Lemen}, {Wuelser}, {Duncan}, {Tarbell},
  {Wolfson}, {Moore}, {Harrison}, {Waltham}, {Lang}, {Davis}, {Eyles},
  {Mapson-Menard}, {Simnett}, {Halain}, {Defise}, {Mazy}, {Rochus}, {Mercier},
  {Ravet}, {Delmotte}, {Auchere}, {Delaboudiniere}, {Bothmer}, {Deutsch},
  {Wang}, {Rich}, {Cooper}, {Stephens}, {Maahs}, {Baugh}, {McMullin}, \&
  {Carter}}]{howard08}
{Howard}, R.~A., Moses, J.~D., Vourlidas, A., Newmark, J.~S.,
Socker, D.~G., Plunkett, S.~P., {\it et~al.}: 2008,
{\it Space Sci. Rev.} {\bf 136}, 67.


\bibitem[{{Jin} {\it et al.}(2009){Jin}, {Ding}, {Chen}, {Fang}, \&
  {Imada}}]{jin09}
{Jin}, M., {Ding}, M.~D., {Chen}, P.~F., {Fang}, C., {Imada}, S.: 2009,
{\apj} {\bf 702}, 27.

\bibitem[{{Kaiser} {\it et al.}(2008){Kaiser}, {Kucera}, {Davila}, {St.~Cyr},
  {Guhathakurta}, \& {Christian}}]{kaiser08}
{Kaiser}, M.~L., {Kucera}, T.~A., {Davila}, J.~M., {St.~Cyr}, O.~C.,
  {Guhathakurta}, M., {Christian}, E.: 2008,
{\it Space Sci. Rev.} {\bf 136}, 5.


\bibitem[{{Kienreich} {\it et al.}(2009){Kienreich}, {Temmer}, \&
  {Veronig}}]{kienreich09}
{Kienreich}, I.~W., {Temmer}, M., {Veronig}, A.~M.: 2009,
{\apjl} {\bf 703}, 118.

\bibitem[{{Kosugi} {\it et al.}(2007){Kosugi}, {Matsuzaki}, {Sakao}, {Shimizu},
  {Sone}, {Tachikawa}, {Hashimoto}, {Minesugi}, {Ohnishi}, {Yamada}, {Tsuneta},
  {Hara}, {Ichimoto}, {Suematsu}, {Shimojo}, {Watanabe}, {Shimada}, {Davis},
  {Hill}, {Owens}, {Title}, {Culhane}, {Harra}, {Doschek}, \&
  {Golub}}]{kosugi07}
{Kosugi}, T., Matsuzaki, K., Sakao, T., Shimizu, T.,
Sone, Y., Tachikawa, S., {\it et~al.}: 2007, {\solphys} {\bf 243}, 3.



\bibitem[{{Luhmann} {\it et al.}(2008){Luhmann}, {Curtis}, {Schroeder}, {McCauley},
  {Lin}, {Larson}, {Bale}, {Sauvaud}, {Aoustin}, {Mewaldt}, {Cummings},
  {Stone}, {Davis}, {Cook}, {Kecman}, {Wiedenbeck}, {von Rosenvinge}, {Acuna},
  {Reichenthal}, {Shuman}, {Wortman}, {Reames}, {Mueller-Mellin}, {Kunow},
  {Mason}, {Walpole}, {Korth}, {Sanderson}, {Russell}, \& {Gosling}}]{luh08}
{Luhmann}, J.~G., Curtis, D.~W., Schroeder, P., McCauley, J.,
Lin, R.~P., Larson, D.~E., {\it et~al.}: 2008,
{\it Space Sci. Rev.} {\bf 136}, 117.

\bibitem[{{Mandrini} {\it et al.}(2007){Mandrini}, {Nakwacki}, {Attrill}, {van
  Driel-Gesztelyi}, {D{\'e}moulin}, {Dasso}, \& {Elliott}}]{mandrini07}
{Mandrini}, C.~H., {Nakwacki}, M.~S., {Attrill}, G., {van Driel-Gesztelyi}, L.,
  {D{\'e}moulin}, P., {Dasso}, S., {Elliott}, H.: 2007,
{\solphys} {\bf 244}, 25.


\bibitem[{{Mari{\v c}i{\'c}} {\it et al.}(2007){Mari{\v c}i{\'c}}, {Vr{\v s}nak}, {Stanger}, {Veronig}, {Temmer}, \& {Ro{\v s}a}}]{maricic07}
{Mari{\v c}i{\'c}}, D., Vr{\v s}nak, B., Stanger, A.~L., Veronig, A.~M.,
Temmer, M., Ro{\v s}a, D.: 2007, {\solphys} {\bf 241}, 99.


\bibitem[{{McIntosh} {\it et al.}(2007){McIntosh}, {Leamon}, {Davey}, \&
  {Wills-Davey}}]{mcintosh07}
{McIntosh}, S.~W., {Leamon}, R.~J., {Davey}, A.~R., {Wills-Davey}, M.~J.: 2007,
{\apj} {\bf 660}, 1653.

\bibitem[M\"ostl {\it et al.}(2011)]{moestl11}
{M\"ostl}, C., {Rollett}, T., {Lugaz}, N., {Farrugia}, C.~J., {Davies}, J.~A., {Temmer}, M., {\it et~al.}: 2007,
{\apj}, {accepted}.


\bibitem[{{Moreton} and {Ramsey}(1960)}]{moreton60}
{Moreton}, G.~E., {Ramsey}, H.~E.: 1960, {\pasp} {\bf 72}, 357.


\bibitem[{{Moses} {\it et al.}(1997){Moses}, {Clette}, {Delaboudini{\`e}re},
  {Artzner}, {Bougnet}, {Brunaud}, {Carabetian}, {Gabriel}, {Hochedez},
  {Millier}, {Song}, {Au}, {Dere}, {Howard}, {Kreplin}, {Michels}, {Defise},
  {Jamar}, {Rochus}, {Chauvineau}, {Marioge}, {Catura}, {Lemen}, {Shing},
  {Stern}, {Gurman}, {Neupert}, {Newmark}, {Thompson}, {Maucherat},
  {Portier-Fozzani}, {Berghmans}, {Cugnon}, {van Dessel}, \&
  {Gabryl}}]{moses97}
{Moses}, D., Clette, F., Delaboudini\`ere, J.-P., Artzner, G.~E.,
Bougnet, M., Brunaud, J., {\it et~al.}: 1997, {\solphys} {\bf 175}, 571.


\bibitem[{{Neupert}(1968)}]{neup68}
{Neupert}, W.~M.: 1968, {\apjl} {\bf 153}, 59.

\bibitem[{{Patsourakos} and {Vourlidas}(2009)}]{patsourakos09}
{Patsourakos}, S., {Vourlidas}, A.: 2009, {\apjl} {\bf 700}, 182.


\bibitem[{{Reinard} and {Biesecker}(2008)}]{reinard08}
{Reinard}, A.~A., {Biesecker}, D.~A.: 2008, {\apj} {\bf 674}, 576.

\bibitem[{{Sterling} and {Hudson}(1997)}]{sterling97}
{Sterling}, A.~C., {Hudson}, H.~S.: 1997, {\apjl} {\bf 491}, 55.

\bibitem[{{Temmer} {\it et al.}(2010){Temmer}, {Veronig}, {Kontar}, {Krucker}, \& {Vr{\v s}nak}}]{temmer08}
{Temmer}, M., Veronig, A.~M., Kontar, E.~P., Krucker, S.,
Vr{\v s}nak, B.: 2010, {\apj} {\bf 712}, 1410.

\bibitem[{{Temmer} {\it et al.}(2008){Temmer}, {Veronig}, {Vr{\v s}nak}, {Ryb{\'a}k}, {G{\"o}m{\"o}ry}, \& {Stoiser}}]{temmer10}
{Temmer}, M., Veronig, A.~M., Vr{\v s}nak, B., Ryb{\'a}k, J.,
G\"om\"ory, P., Stoiser, S., {Mari{\v c}i{\'c}}, D.: 2008, {\apjl} {\bf 673},
95.


\bibitem[{{Thompson} {\it et al.}(2000){Thompson}, {Cliver}, {Nitta},
  {Delann{\'e}e}, \& {Delaboudini{\`e}re}}]{thomson00}
{Thompson}, B.~J., {Cliver}, E.~W., {Nitta}, N., {Delann{\'e}e}, C.,
  {Delaboudini{\`e}re}, J.: 2000, {\grl} {\bf 27}, 1431.

\bibitem[{{Thompson} {\it et al.}(1998){Thompson}, {Plunkett}, {Gurman}, {Newmark},
  {St.~Cyr}, \& {Michels}}]{thompson98}
{Thompson}, B.~J., {Plunkett}, S.~P., {Gurman}, J.~B., {Newmark}, J.~S.,
  {St.~Cyr}, O.~C., {Michels}, D.~J.: 1998, {\grl} {\bf 25}, 2465.

\bibitem[{{T\"or\"ok} and {Kliem}(2003)}]{torok03}
{T\"or\"ok}, T., {Kliem}, B.: 2003, {\aap} {\bf 406}, 1043.



\bibitem[{{Veronig} {\it et al.}(2005){Veronig}, {Brown}, {Dennis}, {Schwartz},
  {Sui}, \& {Tolbert}}]{veronig05}
{Veronig}, A.~M., {Brown}, J.~C., {Dennis}, B.~R., {Schwartz}, R.~A., {Sui},
  L., {Tolbert}, A.~K.: 2005, {\apj} {\bf 621}, 482.



\bibitem[{{Vr{\v s}nak} {\it et al.}(2007){Vr{\v s}nak}, {Mari{\v c}i{\'c}}, {Stanger}, {Veronig}, {Temmer}, \& {Ro{\v s}a}}]{vrsnak07}
Vr{\v s}nak, B., {Mari{\v c}i{\'c}}, D., Stanger, A.~L., Veronig, A.~M.,
Temmer, M., Ro{\v s}a, D.: 2007, {\solphys} {\bf 241}, 85.


\bibitem[{{Webb} {\it et al.}(2000){Webb}, {Lepping}, {Burlaga}, {DeForest},
  {Larson}, {Martin}, {Plunkett}, \& {Rust}}]{webb00}
{Webb}, D.~F., {Lepping}, R.~P., {Burlaga}, L.~F., {DeForest}, C.~E., {Larson},
  D.~E., {Martin}, S.~F., {Plunkett}, S.~P., {Rust}, D.~M.: 2000,
{\jgr} {\bf 105}, 27251.


\bibitem[{{Wuelser} {\it et al.}(2004){Wuelser}, {Lemen}, {Tarbell}, {Wolfson},
  {Cannon}, {Carpenter}, {Duncan}, {Gradwohl}, {Meyer}, {Moore}, {Navarro},
  {Pearson}, {Rossi}, {Springer}, {Howard}, {Moses}, {Newmark},
  {Delaboudiniere}, {Artzner}, {Auchere}, {Bougnet}, {Bouyries}, {Bridou},
  {Clotaire}, {Colas}, {Delmotte}, {Jerome}, {Lamare}, {Mercier}, {Mullot},
  {Ravet}, {Song}, {Bothmer}, \& {Deutsch}}]{wuelser04}
{Wuelser}, J., Lemen, J.~R., Tarbell, T.~D., Wolfson, C.~J.,
Cannon, J.~C., Carpenter, B.~A., {\it et~al.}: 2004,
In: Fineschi, S., Gummin, M.~A. (eds.),
{\it Telescopes and Instrumentation for Solar Astrophysics},
{\it Proc. SPIE} {\bf 5171}, 111.

\bibitem[{{Zarro} {\it et al.}(1999){Zarro}, {Sterling}, {Thompson}, {Hudson}, \&
  {Nitta}}]{zarro99}
{Zarro}, D.~M., {Sterling}, A.~C., {Thompson}, B.~J., {Hudson}, H.~S.,
{Nitta}, N.: 1999, {\apjl} {\bf 520}, 139.

\bibitem[{{Zhang} {\it et al.}(2001){Zhang}, {Dere}, {Howard}, {Kundu}, \& {White}}]{zhang01}
{Zhang}, J., Dere, K.~P., Howard, R.~A., Kundu, M.~R., White, S.~M.: 2001,
{\apj} {\bf 559}, 452.

\bibitem[{{Zhang} {\it et al.}(2006){Zhang}, {Baumjohann}, {Delva}, {Auster},
  {Balogh}, {Russell}, {Barabash}, {Balikhin}, {Berghofer}, {Biernat},
  {Lammer}, {Lichtenegger}, {Magnes}, {Nakamura}, {Penz}, {Schwingenschuh},
  {V{\"o}r{\"o}s}, {Zambelli}, {Fornacon}, {Glassmeier}, {Richter}, {Carr},
  {Kudela}, {Shi}, {Zhao}, {Motschmann}, \& {Lebreton}}]{zha06}
{Zhang}, T.~L., Baumjohann, W., Delva, M., Auster, H.-U.,
Balogh, A., Russell, C.~T.; 2006, {\planss} {\bf 54}, 1336.

\bibitem[{{Zhukov} and {Auch{\`e}re}(2004)}]{zhukov04}
{Zhukov}, A.~N., {Auch{\`e}re}, F.: 2004, {\aap} {\bf 427}, 705.


\end{thebibliography}

\end{article}
\end{document}